\documentclass[useAMS,usenatbib,referee]{mn2e}
\usepackage{graphicx}
\usepackage{epstopdf}
\usepackage[figuresright]{rotating}
\usepackage{subfigure,color}
\usepackage{longtable}
\usepackage{multicol}
\usepackage{supertabular}
\usepackage{lscape}
\usepackage{xspace}
\usepackage{natbib}

\newcommand{\ii}{\'\i}
\newcommand{\ion}[2]{{\textrm{#1}}\,{\textrm{\sc #2}}}

\def\mnras{MNRAS}%
\def\pasp{PASP}%
\title{Interaction effects on galaxy pairs with Gemini/GMOS-\,II: Oxygen abundance gradients}

\author[Rosa et al.]
{D.~A.~Rosa$^1$\thanks{E-mail:deiserosa@univap.br}, O.~L.~Dors Jr.$^1$, A.~C.~Krabbe$^{1}$, G. F. H\"agele$^{2,3}$, 
M.~V. Cardaci$^{2,3}$, \newauthor M.~G.~Pastoriza$^4$,  I. Rodrigues$^1$, C.~Winge$^5$.  \\
$^1$ Universidade do Vale do Para\'iba, Av. Shishima Hifumi, 2911, Cep
12244-000, S\~ao Jos\'e dos Campos, SP, Brazil\\
$^2$ Instituto de Astrof\'isica de La Plata (CONICET La Plata--UNLP), Argentina. \\
$^3$ Facultad de Ciencias Astron\'omicas y Geof\'isicas, Universidad Nacional de La Plata, Paseo del Bosque s/n, 1900 La Plata, Argentina \\
$^4$ Instituto de F\ii sica, Universidade Federal do Rio Grande do Sul, Av.~Bento Gon\c{c}alves, 9500,  Cep 91359-050, Porto Alegre, RS, Brazil\\
$^5$ Gemini Observatory, c/o AURA Inc., Casilla 603, La Serena, Chile \\
}
\begin{document}

\date{Accepted -. Received -.}

\pagerange{\pageref{firstpage}--\pageref{lastpage}} \pubyear{2014}

\maketitle

\label{firstpage}


\begin{abstract}

In this paper we derived oxygen abundance gradients from 
 \ion{H}{ii} regions located in eleven galaxies  in eight systems of close pairs. 
 Long-slit spectra  in the range 4400-7300 \AA{} were obtained  with the Gemini Multi-Object Spectrograph at Gemini South (GMOS). 
Spatial profiles of oxygen abundance  in the gaseous phase along galaxy disks
were obtained  using  calibrations based on strong emission-lines ($N2$ and $O3N2$). 
We found oxygen gradients  significantly flatter for all the studied
  galaxies than those in typical isolated spiral galaxies. 
 Four objects in our sample,  AM\,1219A, AM\,1256B, AM\, 2030A and
  AM\,2030B, show a clear break in the
oxygen abundance at galactocentric radius $R/R_{25}$ between 0.2 and
0.5. 
For AM\,1219A and AM\,1256B we found negative slopes for
the inner gradients, and for AM\,2030B we found a positive one. In all
these three cases they show a flatter behaviour to the outskirts of the galaxies. 
For AM\,2030A, we found a positive-slope outer gradient while the inner one is
almost compatible with a flat behaviour.
A decrease of star formation efficiency in the zone that
corresponds to the oxygen abundance gradient break for AM\,1219A and
AM\,2030B was found. For the former, a minimum
in the estimated metallicities was found very close to the break zone that could 
be associated with a corotation radius. 
On the other hand, AM\,1256B and AM\,2030A, present a
SFR maximum but not an extreme oxygen abundance value.
All the four interacting systems that show oxygen gradient breakes the
extreme SFR values are located very close to break zones.
\ion{H}{ii} regions located in 
close pairs of galaxies 
follow the same relation between the
ionization parameter and the oxygen abundance as those regions in
isolated galaxies.

\end{abstract}


\section{Introduction}

 The  study of the chemical evolution of galaxies, both isolated and interacting, 
 play an important role  to understand  the  formation of these objects, its stellar formation history
and the evolution of the Universe.

In general, for almost all disk isolated galaxies   a   negative
oxygen gradient is derived, such as our Galaxy  \citep{bragaglia08, magrini09, yong14, pedicelli09, andrievsky02, andrievsky04, luck03, lemasle13,  
esteban13, vilchez96, costa04, maciel09}. This negative gradient is naturally explained by models which assume 
the  growth in the inside-out scenario of galaxies \citep{portinari99, boissier00, molla05}, where galaxies begin
to form their inner regions before the outer ones, as confirmed by  stellar populations studies of spiral galaxies
\citep{bell00, mac04, pohlen06, munoz07} and by very deep photometric studies of galaxies at high redshifts
\citep{trujillo04, barden05}.

 The  oxygen gradients can be flattened or modified by the presence of 
gas flows along the galactic disk. Basically,  these gas flows could be arise due to two mechanisms. 
In isolated galaxies, hydro-dynamical simulations predict that 
 the bars may produce a falling of gas into the central regions \citep{athanassoula92, friedli94} which have been
 confirmed by observational studies   \citep{zaritsky94, martin94}. 
The  second mechanism occur in interacting galaxies or close pairs, where interaction-induced gas flows from the outer parts to the centre of 
 each component  \citep{dalcanton, toomre72}  seem  to modify  the metallicity in  galactic disks. Therefore, 
 the metallicity gradients of  interacting galaxies 
or  galaxies that have had  an interaction in the past are shallower \citep{sanchez14, miralles14, rupke10a,  bresolin09}  
than the ones derived for isolated galaxies (\citealt{sanchez12a, rupke10a}).
 In fact, \citet{krabbe08,krabbe11} combining long-slit spectroscopy data for the
  interacting pairs AM\,2306-721 and  AM\,2322-821 with grids of photoionization models found  
shallower metallicity gradients than the ones in isolated spiral galaxies.  However,
two  works  carried out  the first systematic investigations about metallicity gradients  in interacting galaxies.
(a) \citet{lisa10} determined the metallicity gradients for eight galaxies in
close pairs and found  them
shallower than gradients in isolated spiral galaxies.  (b) \citet{sanchez14}, using data obtained from the CALIFA survey, 
 found  that galaxies with evidence of interactions and/or clear merging systems present a significant shallower gradient.

Furthermore, the gas motions produced by the interactions  also  induced star formation
  along the disk of the galaxies involved \citep{alonso12}, and this  burst of star formation may be 
 associated with a flatter metallicity gradient \citep{lisa10}. For example,  \citet{chien07} determined the 
 oxygen abundance of 12 young star clusters in the merging galaxy pair NGC\,4676.
 These authors found a nearly flat oxygen distribution along the northern tail of this object,
  suggesting efficient gas mixing (see also \citealt{bastian09, trancho07}). 
 Recently, \citet{scudder12}, using a large sample
of galaxy pairs taken from the Sloan Digital Sky Survey Data Release 7, found 
that galaxies in pairs show a star formation rate (SFR) about 60\% higher
than the one in non-pair galaxies (see also \citealt{nikolic04, lambas03, barton00}).
Additional analysis of these data by \citet{sara13}, who investigated the effects of galaxy mergers throughout the interaction sequence, 
re\-ve\-led an enhancement of the average central SFR by a factor of
about 3.5  in relation to the one in objects with no close companion. \citet{sara13}
also found a stronger deficit in the gas phase metallicity in the
post-merger sample than in   closest pairs  (see also \citealt{alonso12, barton00, bernloehr93, bergvall03, lambas03,
dimatteo08, patton11, woods10, mihos10}). 

Although recent efforts in the direction to understand the effects of  interactions on chemical evolution of
galaxies have been done (\citealt{krabbe08, lisa10, krabbe11,bresolin12, flores14}),
 the number of galaxies in close pairs for which the  metallicity have been estimated along their galactic disks is insufficient for a statistical analysis.
 To increase the number of determinations of metallicity gradients in galaxy pairs  producing a better knowledge  of the several
phenomena that  arise during the interactions is the main goal of this
paper. 

In a previous work (\citealt{krabbe14}, hereafter Paper I), we presented  an observational study of the impact of the interactions on the electron density 
of \ion{H}{ii} regions located in seven systems of interacting galaxies. We found that the electron density estimates obtained in our sample
are  systematically higher than those derived for isolated galaxies.  In the
present paper, we mainly use these data to estimate the metallicity
gradients along the disks of eight galaxy pairs. This work is organized as follows. In Section~\ref{obser_data}
we summarize the observations and data reduction. In Sect.~\ref{determinations}
  the method to compute the  metallicity of the gas phase of our sample
is described.  Results and discussion are
presented in Sections~\ref{results} and \ref{discussion}, respectively.
The conclusions of the outcomes are given in Section~\ref{conc}.

\begin{figure} 
\centering
\includegraphics*[angle=270, width=0.5\textwidth]{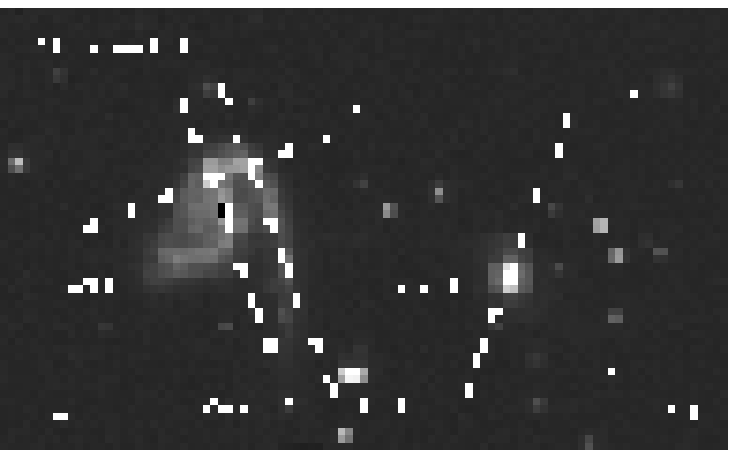}
\includegraphics*[width=0.5\textwidth]{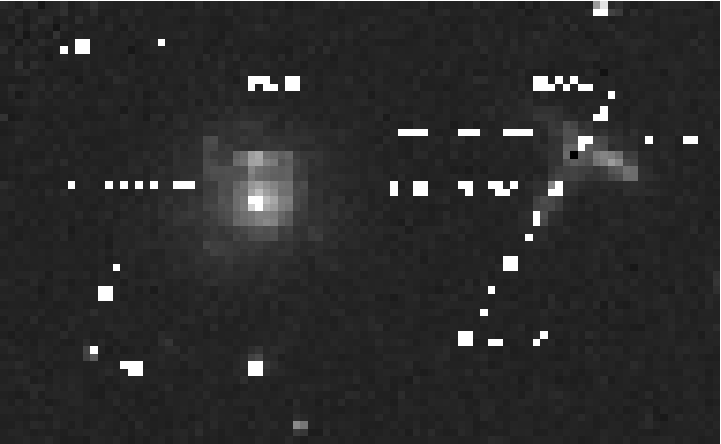}
\includegraphics*[width=0.5\textwidth]{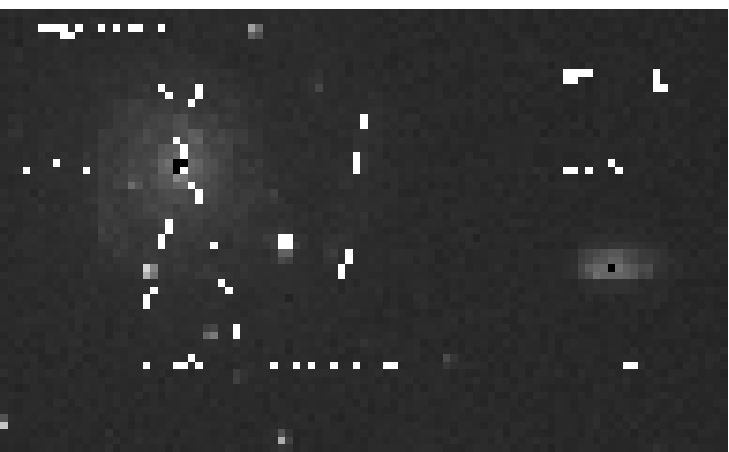}
\caption{Slit positions for AM\,1219-430, AM\,2030-303 and AM\,2322-821 systems superimposed on the GMOS-S r$\arcmin$ acquisition images.}
\label{fendas_1}
\end{figure}


\section{Observational data}
\label{obser_data}

  We have selected eight close pairs systems from \citet{ferreiro04} to
study the effects of minor mergers on gradient abundances of the individual galaxies.
Objects with  mass ratio in the range of 
$0.04 < M_{secondary}/M_{primary} < 0.2$, apparent B magnitude  higher than 18,
 redshift in the range  $  0.01 \: \la \: z \: \la \:  0.06$, and classified as close interacting pairs  were selected.

Long-slit spectroscopic data of the galaxy systems AM1054-325,
AM\,1219-430,  AM\,1256-433,  AM\,2030-303,  AM\,2058-381, AM\,2229-735,
AM\,2306-721, and  AM\,2322-821 were obtained with the Gemini Multi-Object Spectrograph
  (GMOS) attached to the 8\,m Gemini South telescope.  Spectra in the range 4400\AA-7300\AA  
\, were acquired with the B600 grating, a  slit width of 1 arcsec and a
spectral 
resolution of $\sim 5.5$ \AA. Except for AM\,2030-303, detailed
information on the galaxy systems observed,  containing most of the slit positions
for each system and a complete description of the data reduction  were
presented in Paper I, and are not reproduced here.

AM\,2030-303 is the only object in our sample that was not included in Paper I
and  its  main information is presented in Table~\ref{tabw}. 
For the systems AM\,1256-433 and AM\,2058-381 one more slit
position than the ones presented in Paper I is considered in the present work
(PAs 70\degr and 28\degr, respectively).
It is important to note that for some systems of galaxies (e.g. AM\,2306-721 and AM\,2322-821, see 
\citealt{krabbe08, krabbe11}) we had spectra in the range of 
about 3400\AA-7300\AA. However, in order to obtain homogeneous metallicity determination,
we  restricted  the analysis to the same wavelength range (i.e 4400\AA-7300\AA). In Figure \ref{fendas_1}
the slit positions for these three systems are shown superimposed on the
GMOS-S r$\arcmin$  acquisition image (see Figure 1 of Paper I for the
slit positions of the other five objects). These data were not included in
Paper I due to the low signal-to-noise ratio (S/N) of the [\ion{S}{ii}]$\lambda$6716,$\lambda$6731 emission lines
needed to perform the electron density estimations.
The observed spectra comprise the flux contained in an aperture of
1 arcsec $\times$ 1.152 arcsec which,  considering  a spatially flat  cosmology with
 $H_{0}$\,=\,71 $ \rm km\:s^{-1} Mpc^{-1}$, $\Omega_{m}=0.270$,  $\Omega_{\rm vac}=0.730$  \citep{2006PASP..118.1711W} and
the distances to the systems of our sample,   corresponds to apertures between about 200 and
1100 pc on the plane of the galaxies.
Therefore, the physical properties derived from these spectra represent the ones of a complex
of \ion{H}{ii} regions. In Table~\ref{tabs} we present the nuclear
separation between  the components of the 
galaxy pairs,   the galactocentric distances 
given in units of $R/R_{25}$, where $R_{25}$ is the B-band isophote at a surface
brightness of 25 mag $\rm arcsec^{-2}$, the inclination angle ($i$)
of each galaxy analysed, and the references from which the information was taken.
  The inclination of each galaxy with respect to the plane of the
sky was computed as $\cos (i) = b/a$, where $a$ and $b$ are the major and
minor semi-axes of the galaxy, respectively. The $a$ and $b$ values  as well as the position angle of the major axis of each
galaxy were obtained from the  Gemini  acquisition images in the r filter,
using a simple isophotal fitting with the {\sc iraf stsdas.ellipse} task.
The same procedure was used in \citet{krabbe11}.

\begin{table*}
\begin{footnotesize}
\caption{Morphological type, right ascension, declination,  radial velocity, magnitude, and  cross-identifications for AM\,2030-303.} 
\label{tabw}
\begin{tabular}{@{}llllrrl@{}}
\hline
\noalign{\smallskip}
ID         & Morphology   &$\alpha$(2000) & $\delta$(2000)& $cz\,(\rm km/s)$ & $m_{\rm B}$\,(\rm mag)  & Others names \\
\noalign{\smallskip}
\hline
\noalign{\smallskip}
AM\,2030-303 & SA?   [2]     & 20 ~33  ~56.3    & $-$30  ~22 ~41       &  12\,323 [2]    &15.25 [1]   & ESO 463-IG 003 NED01\\
              & G Trpl   [1]   &20 ~33  ~59.7    & $-$30  ~22 ~29      &  12\,465 [2]    &17.80 [1]   & ESO 463-IG 003 NED02\\
              & G Trpl    [1]  &20 ~33  ~59.7    & $-$30  ~22 ~23      &  12\,474 [2]    &21.39 [1]  & ESO 463-IG 003 NED03\\
\hline
\noalign{\smallskip}
\end{tabular}
\begin{minipage}[c]{1\textwidth}
{\it References:} [1] \citet{ferreiro04}; [2] \citet{donzelli97};
{\it Conventions:} $\alpha$, $\delta$: Equatorial Coordinates.\\
\end{minipage}
\end{footnotesize}
\end{table*}

\begin{table*}
\begin{center}
\caption{Nuclear separation (NS) between the galaxies of the pairs, galactocentric distance
 with surface brightness of 25 mag $\rm arcsec^{-2}$
 ($R_{25}$), inclination angle ($i$), and  the radial velocity ($cz$) of the objects in our sample.}
\label{tabs}
\begin{tabular}{@{}lccrrl@{}}
\hline
\noalign{\smallskip}
ID                   &   &NS\,(kpc)& \textit{R}$_{25}$ (kpc)& \textit{i}\,(\degr)& c$z$($\rm km\:s^{-1}$)\\
\hline
AM\,1054-325  & A  &-~~~~   &6.98\,[3]& 62 [2] & 3\,788 \\
                       & B  & 17     & 6.81 [3]&54 [2]    &  3\,850\\
\hline
AM\,1219-430  &  A  &-~~~~   &15.3  [6]      & 50 [6] & 6\,957  \\
                       &  B  & 33.7  & 6.2  [6]     &-~~~~      & 6\,879  \\
\hline		       
AM\,1256-433  &  A  & -~~~~   &-~~~~     &36 [2]   &  9\,215\\
                      &      &-~~~~   & -~~~~    &33 [2]     & 9\,183 \\
                      &  B  & 91.6   &24.32 [4]&77 [4]        & 9\,014\\
\hline		      
AM\,2030-303  &  A   &-~~~~   & 17.4 [4]&23 [4] &  12\,323\\
                       &     & -~~~~ &-~~~~     &-~~~~ &   12\,465\\
                       &  B   &40.5   & 13.5 [4]&35 [4] & 12\,474\\
\hline		       
AM\,2058-381   &  A  & -~~~~  & 34.3 [4]&68 [4] & 12\,383\\
                        & B    & 44     & 16.7 [4]&57 [4] & 12\,460\\
\hline			
AM\,2229-735  & A    & -~~~~   &26.1 [4]& 60 [4] & 17\,535\\
                       & B   & 24.5   &22.5 [4] & 48 [4] & 17\,342\\
\hline		       
AM\,2306-721  &  A    & -~~~~   &24.3 [4]&56 [5] & 8\,919 \\
                       &  B    & 52.6  &15.6 [4]& 60 [5]   &  8\,669  \\
\hline		       
AM\,2322-821  &   A   & -~~~~   &13.5 [4]&20 [1] & 3 680  \\
                       &   B   & 33.7   &4.2 [4] &63 [1]    &  3 376 \\
\hline
\noalign{\smallskip}
\end{tabular}
\begin{minipage}[c]{2\columnwidth}
  References  [1] \citet{krabbe11}; [2] \citet{paturel03};  
  [3] \citet{paturel91};  [4] \citet{ferreiro08}, 
[5] \citet{krabbe08}, \newline [6] \citet{jose13}.

\end{minipage}
\end{center}
\end{table*}

 It can be seen in Tables~\ref{tabw} and \ref{tabs} that some
galaxies of our sample have large inclinations and it could  affect 
the derived abundance gradients. In fact, as pointed out by \citet{sanchez12a},  face-on
galaxies are more suitable to study the spatial distribution of the
properties of \ion{H}{ii} regions.  For example, if we assume
an inclination angle $i$ for a given galaxy larger than the real one, the abundance gradient
derived would be steeper than the one obtained with the right $i$ value. However, this effect is critical for 
isolated spiral galaxies  which have a clear (or steep) abundance gradients
and  it is not so important  for objects with shallow gradients, such as interacting galaxies.

\begin{figure*}
\centering
\includegraphics*[angle=-90,width=\textwidth]{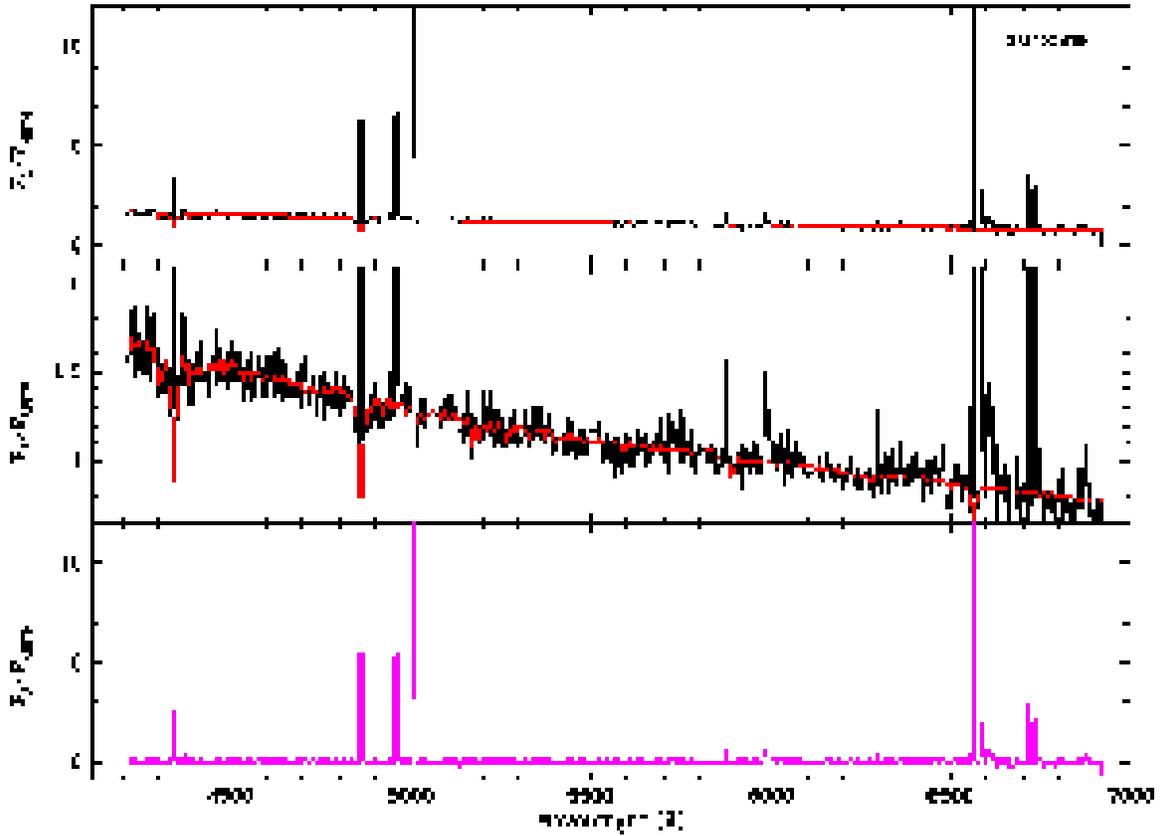}
\caption{Stellar population synthesis for the brightest region of AM\,1256B along the PA=325\degr. 
Top panel: reddening-corrected spectrum (black line) and the synthesized
spectrum (red line). Middle panel: y-axis zoom of the top panel. Bottom
panel: pure emission spectrum obtained as the difference between both
  spectra in top panel.}
\label{sintese_001}
\end{figure*}

 To obtain the nebular spectra not contaminated by the stellar population contribution, we 
 use the stellar population synthesis code STARLIGHT \citep{cid05} following the methodology
 presented in \citet{krabbe11}. Detailed analysis of the  stellar population for the sample
 of objects will be presented in a future work (Rosa et al.\  in preparation). Once the stellar population contribution has been determined, the
underlying absorption line spectrum  was subtracted from the observed spectra. In Fig.~\ref{sintese_001} the spectrum of  the region with the 
 highest brightness   of AM\,1256-433 along the PA=325\degr, the synthesized spectrum and 
 the pure emission spectrum corrected for reddening are shown.  The   intensities   
 of the emission-lines H$\beta$, [\ion{O}{iii}]$\lambda$5007,
H$\alpha$, [\ion{N}{ii}]$\lambda$6584, and
[\ion{S}{ii}]$\lambda\,$6716,$\lambda$\,6731 were  obtained from the
pure nebular spectrum of each aperture  using Gaussian line profile fitting. 

We used the {\sc IRAF}\footnote{Image Reduction and  
Analysis Facility, distributed by NOAO, operated by AURA, Inc., under
agreement with NSF.} {\sc splot} routine to fit the lines, with the associated error given as  $\sigma^{2} =\sigma_{cont}^{2} + \sigma_{line}^{2}$,
where  $\sigma_{cont}$ and $\sigma_{line}$ are the continuum  rms and the Poisson error of the line flux, respectively.
The residual extinction associated with the gaseous component for
each spatial bin was calculated comparing the observed value of H$\alpha$/H$\beta$ ratio to the theoretical value  2.86 obtained by 
 \citet{hummer87}   for an electron temperature of 10\,000 K and an electron
density of 100 $\rm cm^{-3}$.   This value for the electron density is in the range   
of   mean electron density values  ($24 \: \la \: N_{\rm e} \: \la \: 532 \rm \: cm^{-3}$)  found for interacting galaxies in Paper I.
  The correction for foreground dust was done using
the reddening law  given by  \citet{cardelli89}, assuming the specific attenuation  $R_{\rm V} = 3.1$.
We considered only emission-line  measurements  whose S/N was higher than 8.
The galactocentric distance in relation to $R_{25}$, the flux of H$\beta$, the extinction coefficient
$C(\rm H\beta$), and the   emission-line intensities  normalized to the
flux of H$\beta$  for the regions
considered in the systems are presented in Table~\ref{tcorr1}. 

\begin{table*}
\caption{Specific parameters and intensity of emission lines corrected by
  reddening (relative to H$\beta$=\,100). Only the  AM\,1054B galaxy is shown here. The total list of galaxies is available in electronic form.}
\begin{tabular}{|lccccccc}
\hline
$R/R_{25}$           &     $\log [F (\rm H\beta)]^{*}$              &  $C(\rm H\beta$)   &  [O{\sc iii}]$\lambda$5007   &     H$\alpha$         &   [N{\sc ii}]$\lambda$\,6584 &   [S{\sc ii}]$\lambda$\,6716 & [S{\sc ii}]$\lambda$\,6731 \\    
 \hline  
                            &     \multicolumn{7}{c}{AM\,1054B} \\
\noalign{\smallskip}			    
0.00                      &      $-$14.51                   &0.40                     & 152$\pm$27                   & 278$\pm$9       &      112$\pm$6                &        24$\pm$10                &  27$\pm$7 \\ 
0.04SW                 &     $-$14.97                    &0.26                    & 154$\pm$22                    & 281$\pm$7       &      115$\pm$5               &         28$\pm$4                &  29$\pm$4 \\ 
0.04NE                 &      $-$14.93                    &0.23                    & 141$\pm$21                    & 282$\pm$9      &       114$\pm$6               &        37$\pm$3                 & 37$\pm$7\\ 
0.08NE                 &      $-$15.82                    &0.08                    & 139$\pm$11                    & 285$\pm$11    &       119$\pm$8               &          ---                              &    ---              \\ 
0.13NE                 &      $-$17.11                    &0.08                     & 113$\pm$19                    & 285$\pm$15   &       119$\pm$11               &          ---                               &    ---              \\   
\hline
\end{tabular}
\begin{minipage}[l]{13.0cm}
 [*] Logarithm of the H$\beta$ observed flux in  erg s$^{-1}$ cm$^{-2}$.

\end{minipage}
\label{tcorr1}
\end{table*}

\section{Determination of the oxygen abundance gradients}
\label{determinations}

Since emission-line sensitive to the electron temperature are not detected in
the spectra of the objects in our sample,
the metallicity of the gas phase, traced by the relative abundance  of the oxygen to the  hydrogen (O/H), 
was estimated using calibrations based on strong emission-lines. 
 
Considering the emission-lines
observed in our sample, it is only possible to use  the intensities of the emission-lines defined as
$N2$=log([\ion{N}{ii}]$\lambda$6584/H$\alpha$)    and  \\
$O3N2$=log[([\ion{O}{iii}]$\lambda$5007/H$\beta$)/([\ion{N}{ii}]$\lambda$6584/H$\alpha$)]
 proposed by \citet{thaisa94} and \citet{alloin79}, respectively, as O/H indicators. We used the relations among these
 indexes and the O/H, calculated using direct estimations of the oxygen electron temperatures ($T_{e}$-method), proposed
 by  \citet{perez09} and given by

\begin{equation}
\label{n2c}
\centering
12+ \log({\rm O/H})=0.79\,\times\,N2\,+\,9.07
 \end{equation}
 and
\begin{equation}
\label{o3n2c}
\centering
12+\log({\rm O/H})=8.74-0.31\,\times\,O3N2.
 \end{equation}
 
\begin{figure*}
 \includegraphics*[angle=270,width=0.49\textwidth]{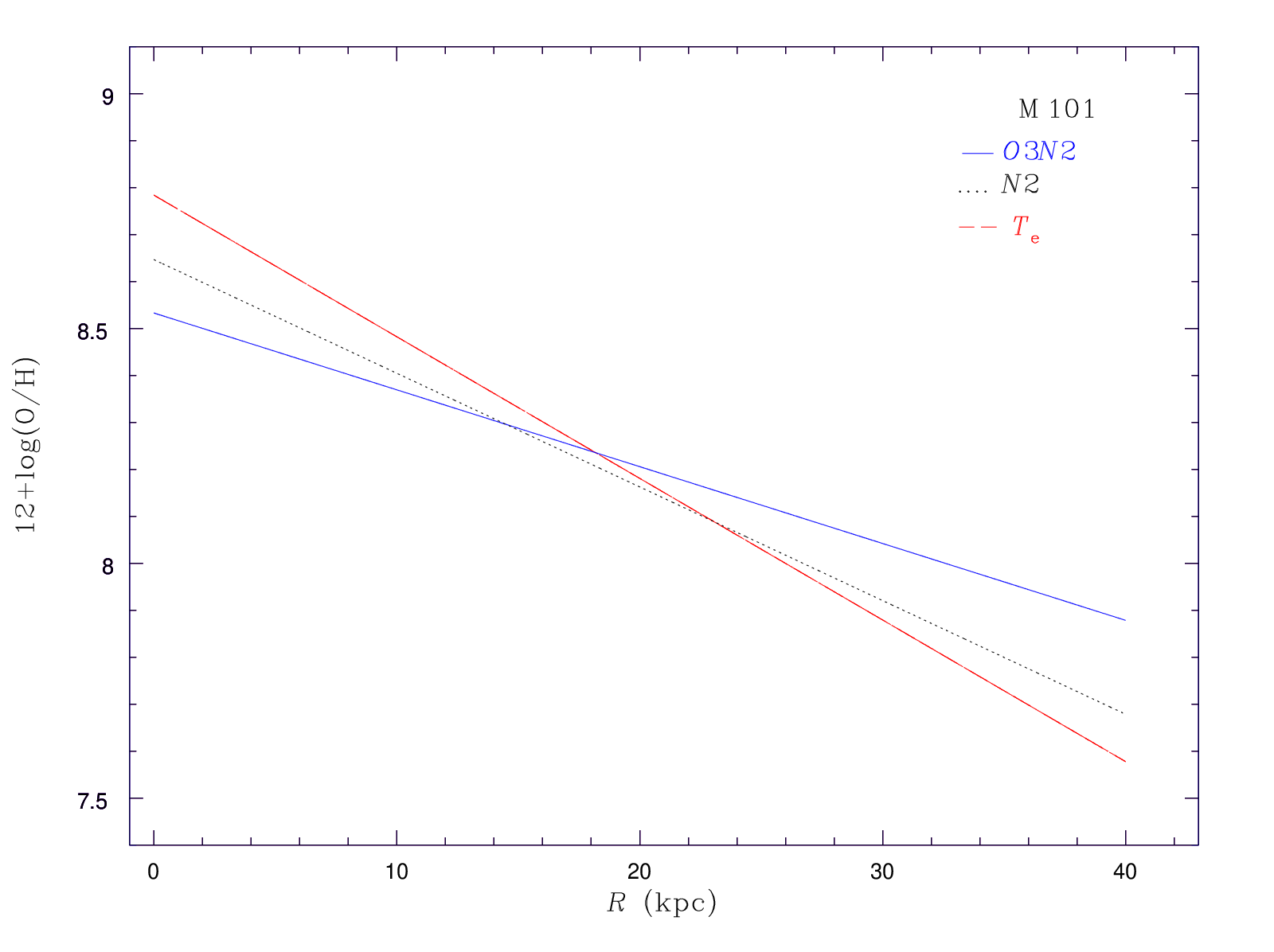}
 \includegraphics*[angle=270,width=0.49\textwidth]{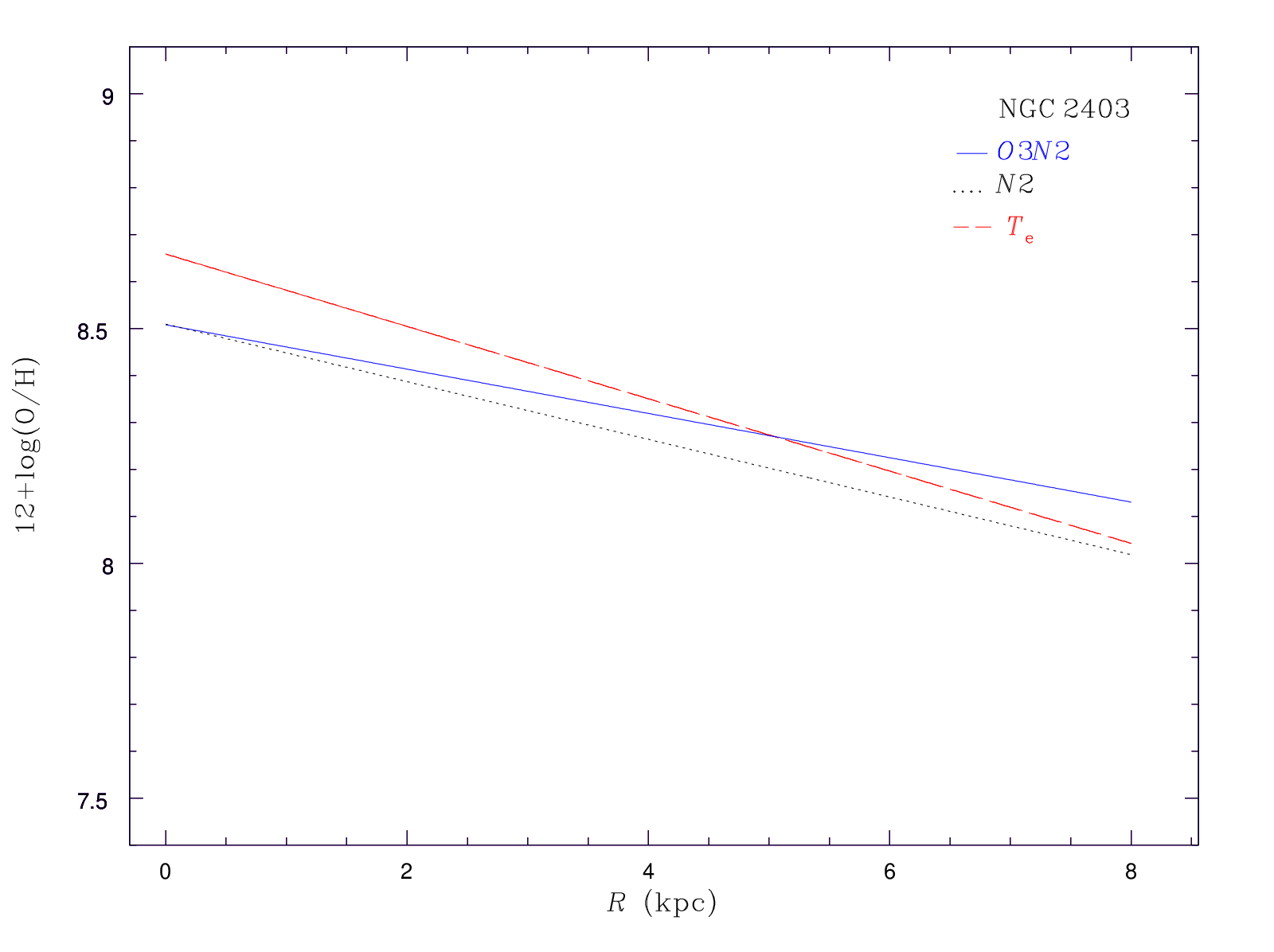}
\caption{Oxygen abundance gradients computed via the $N2$, $O3N2$
    indexes and direct estimations of the oxygen electron temperatures for
  the galaxies M\,101 (left) and NGC\,2403 (right) using data taken from
  \citet{kennicutt03} and  \citet{garnett97}, respectively. }
\label{amostra_1}
\end{figure*}

These calibrations are very similar to the ones proposed by \citet{pettini04}.
 Most recent update of this calibration was done by \citet{marino13}, who used direct oxygen 
abundance measurements obtained from CALIFA survey and other sources from the literature.
 \citet{lopez10}, using  multiwavelength analysis of a sample of  starburst galaxies and data compiled from the literature,
showed that the $N2$ and  $O3N2$ parameters provided   acceptable results for objects with 
12+log(O/H)$>$8.0. \citet{lopez10} also found that empirical calibrations considering  these
indexes give results that are  about  0.15 dex higher than the oxygen
abundances derived  via the $T_{\rm e}$-method.
Despite this difference is similar  to the uncertainties of oxygen abundances derived from the  $T_{\rm e}$-method
 (e.g. \citealt{hagele08, kennicutt03}),
it seems to vary with the regime of metallicity \citep{dors11}.  This can
yield steeper oxygen gradients than the ones from the $T_{\rm e}$-method  or
 erroneous bend in the slope of  abundance gradients \citep{pilyugin03}.   
With the goal to compare O/H gradients derived using the $N2$ and $O3N2$
  indexes with the ones obtained from the $T_{\rm e}$-method,
we  used data of \ion{H}{ii} regions located along the disks of the spiral galaxies  M\,101 e NGC\,2403
obtained by \citet{kennicutt03} and  \citet{garnett97}, respectively. In  Fig.~\ref{amostra_1} these gradients are shown, where we can see
gradients derived from the indexes are somewhat shallower than the ones from $T_{\rm e}$-method.

 \citet{scarano11} and \citet{bresolin12} pointed out that gradients are less 
affected by   uncertainties  in oxygen estimations yielded by calibration of strong emission-lines.
 However, it can be seen in Fig.~\ref{amostra_1} that the value of
 the O/H extrapolated for the 
 central region of the galaxies ($R=0$),  obtained using different methods, can differ by  until 0.4 dex. This  difference is higher
 than the uncertainty attributed to the O/H  estimations using    strong emission-line  calibrations \citep{lisa08}.
  Similar results were found by \citet{bresolin11} comparing the oxygen gradient for NGC\,4258
 using  theoretical calibrations by \citet{mcGaugh91}  and  empirical ones by  \citet{pilyugin05}. The latter
 provides results essentially in consonance  with those obtained from the  $T_{\rm e}$-method (see also \citealt{pilyugin12}).

\section{Results}
\label{res3}

\label{results}
\begin{table*}
\caption{Slope of the oxygen abundance gradient  and the  central value
  derived for the objects in our sample   with no bi-modal gradient behaviour.}
\vspace{0.3cm}
\begin{tabular}{|lccccc}
\hline
Object          
                                &                     \multicolumn{2}{c}{$N2$}                   &       &           \multicolumn{2}{c}{$O3N2$}        \\
                                &     Slope  $[{\rm dex}/(R/R_{25})]$    &  12+log(O/H)$_{\rm Central}$       &       &  Slope  $[{\rm dex}/(R/R_{25})]$    &  12+log(O/H)$_{\rm Central}$   \\
\cline{2-3}
\cline{5-6}	
\noalign{\smallskip}	    
AM\,1054B                &     $+0.11\pm0.04$              &    8.54$\pm0.01$   &       &        $+0.36\pm$0.04            &     8.75$\pm0.06$   \\
                                &                                          &                             &       &                                            &                              \\
AM\,1219B               &     $+0.10\pm0.18$              &   8.89$\pm0.04$   &       &        ---                                 &   ---                         \\ 
                            &                                             &                             &       &                                            &                               \\
AM\,2058A              &    $-0.29\pm0.08$               &    8.79$\pm 0.02$   &       &        $-0.35\pm0.08$               &   8.78$\pm 0.02$   \\
                              &                                            &                             &       &                                               &                             \\
AM\,2229A              &     $+0.03\pm0.09$               &    8.71$\pm 0.03$  &       &        $-0.11\pm0.06$               &   8.70$\pm 0.02$  \\ 
                              &                                            &                              &       &                                              &                           \\
AM\,2306A             &       $-0.40\pm$0.05              &    8.81$\pm0.02$   &       &         $-0.57\pm$0.06               &   8.83$\pm0.02$    \\
                            &                                           &                              &       &                                               &                             \\
AM\,2322A           &       $-0.17\pm0.01$             &     8.79$\pm0.01$   &       &         $-0.18\pm0.02$               &   8.77$\pm 0.01$    \\
AM\,2322B           &       $-0.14\pm0.05$             &     8.57$\pm 0.02$  &       &         $-0.07\pm0.05$               &   8.53$\pm 0.02$   \\
\hline
\end{tabular}
\label{tab33}
\end{table*}
 
In Figs.~\ref{grad_abu_1054B}-\ref{grad_abu_2322} the oxygen abundance determinations along the disks
of the galaxies of our sample, obtained  using Eqs.~\ref{n2c} and \ref{o3n2c},  and  linear regression fits
to these data are presented. In Table~\ref{tab33} the slopes of these  fits
and the  values of 12+log(O/H) central ($R=0$) for  the galaxies which   only 
a global gradient represents well the O/H disk distribution  (i.e.  no bi-modal
gradient behaviour were derived) are presented.
It can be noted that in most cases, shallow gradients
are derived in the interacting galaxies.   The average values of the
 global  gradients calculated for the close pairs in our sample are   
 $-0.10\pm0.19$ $[{\rm dex}/(R/R_{25})]$ and $-0.15\pm0.31$ $[{\rm dex}/(R/R_{25})]$ using the $N2$ and $O3N2$ indexes,
  respectively.  These values are in consonance with
the mean gradient  $-0.25\pm0.02$ $[{\rm dex}/(R/R_{25})]$
derived by \citet{lisa10} and they are shallower than  the mean
metallicity gradient $-0.57\pm0.19$  $[{\rm dex}/(R/R_{25})]$  derived for  11 isolated spiral galaxies  by
\citet{rupke10b}.   
 \citet{sanchez14}, using the CALIFA data survey, presented a study of galaxies with different 
interaction stages in order  to study the  effect on the abundance gradient. This study has a stronger statistical significance  than  the one in previous studies. 
\citet{sanchez14} showed the distribution of slopes of the abundance gradients derived  for the different classes based on the interaction
stages. From this analysis, \citet{sanchez14} found that galaxies with not evidence of interaction have an average
value for the gradient of $-0.11$ dex/$r_{\rm e}$ and objects with evidence for early or advanced interactions
have a slope of $-0.05$ dex/$r_{\rm e}$, being $r_{\rm e}$ the disk effective radius.
This result  confirms the our  findings and the ones  obtained by \citet{lisa10}.
In what follows, the results obtained for each system are discussed separately.

\subsection{AM\,1054-325}
 
  This system is composed by two galaxies, one main galaxy namely  AM\,1054A and other secondary AM\,1054B.
  Using the diagnostic diagram [\ion{O}{iii}]$\lambda$5007/H$\beta$  versus [\ion{O}{i}]$\lambda$6300/H$\alpha$,
  we found (see Paper I) that almost all \ion{H}{ii} regions located in the disk of AM\,1054A  
  have emission lines excited by shock gas. Therefore, abundance determinations was not performed for
  this object  since shocks alter the ionization in a way that the abundance calibrators can not be used due 
to they are calibrated for \ion{H}{ii} regions dominated by photoionization by young stars. 
 
\begin{figure*}
 \includegraphics*[angle=270,width=0.8\textwidth]{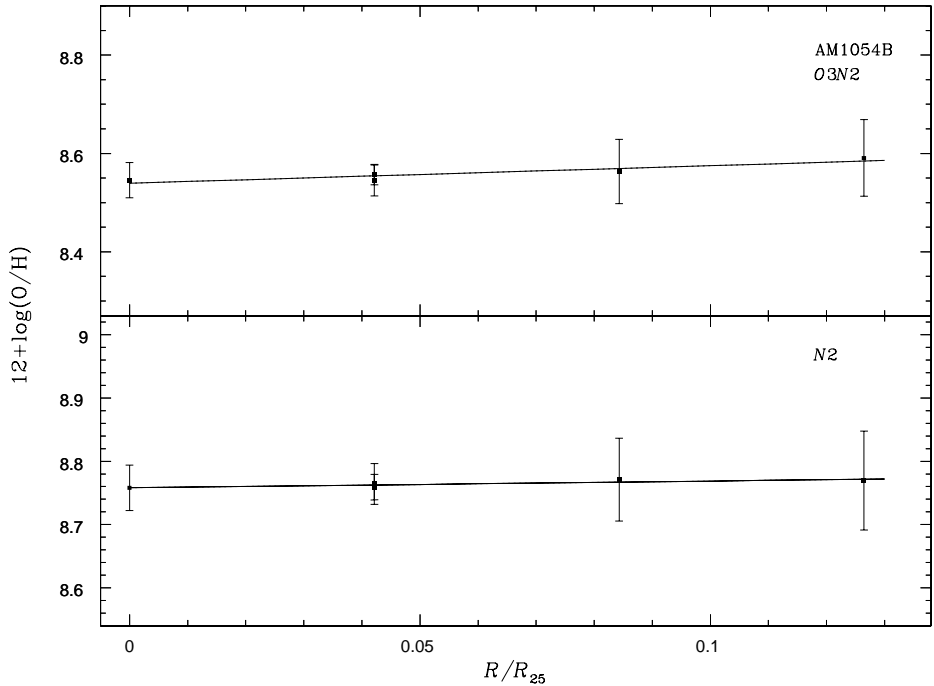}
\caption{Oxygen abundance estimations along the disk of AM\,1054B vs.\ 
  the galactocentric distance given in $R/R_{25}$.
Points represent oxygen estimations obtained via the  $N2$ (lower panel)
  and $O3N2$ (upper panel) indexes. Lines represent a linear
regression fit to our estimations whose coefficients are given in
Table~\ref{tab33}.}
\label{grad_abu_1054B}
\end{figure*}

 In  Fig. \ref{grad_abu_1054B} the O/H distribution versus the galactocentric radius $R$ normalized by 
$R_{25}$ for AM\,1054B is shown. We found gradient slopes of 
\,0.11$\pm$0.04  and 0.36$\pm$0.04  ${\rm dex}/(R/R_{25})$, with  the central 12+log(O/H) value  being  8.54\,($\pm$0.01) and 
  8.75\,($\pm$0.06) dex from the $N2$ and $O3N2$ indexes, respectively. AM\,1054B is the only object in our sample for which both
  estimations of the central oxygen 
  abundances are not in agreement among themselves within the errors (see Table
  \ref{tab33}).  However, as can be seen in Fig. \ref{grad_abu_1054B}, the slopes were obtained using  few points. 
   Hence, the gradient determination is highly uncertain, although the current data indicate a flat O/H distribution.

\subsection{AM\,1219-430}

This system is composed by the  main  galaxy   AM\,1219A and a secondary  galaxy,  AM\,1219B.
Since it was not possible to measure the  [\ion{O}{iii}]$\lambda$\,5007
emission line with enough S/N in our spectrum of AM\,1219B, for
this object O/H was estimated only using $N2$. We corrected by inclination the
galactocentric distances for the main galaxy AM\,1219A    
considering \textit{i}=50$^{\circ}$.

\begin{figure*}
 \includegraphics*[angle=270,width=0.8\textwidth]{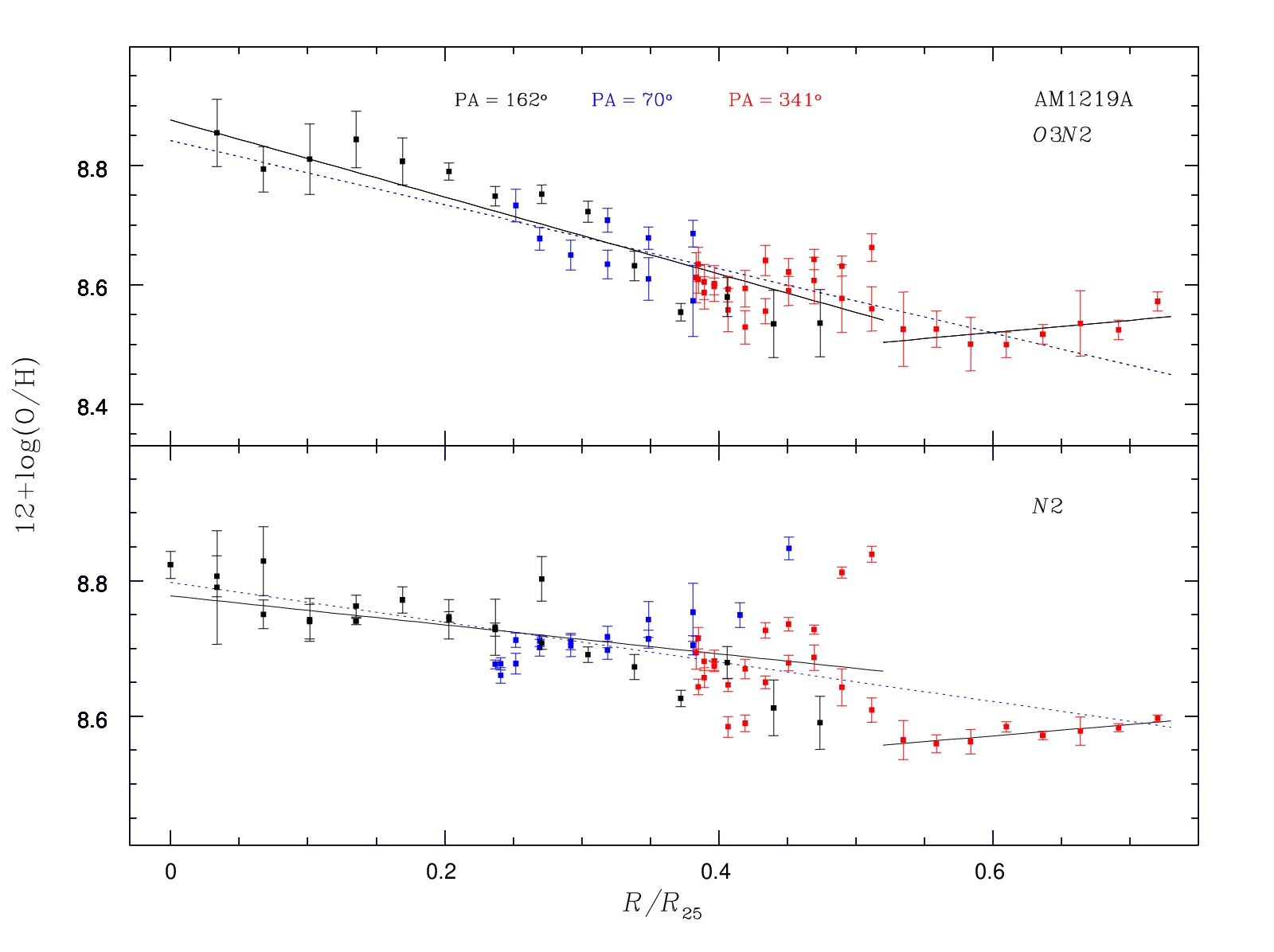} 
 \includegraphics*[angle=270,width=0.8\textwidth]{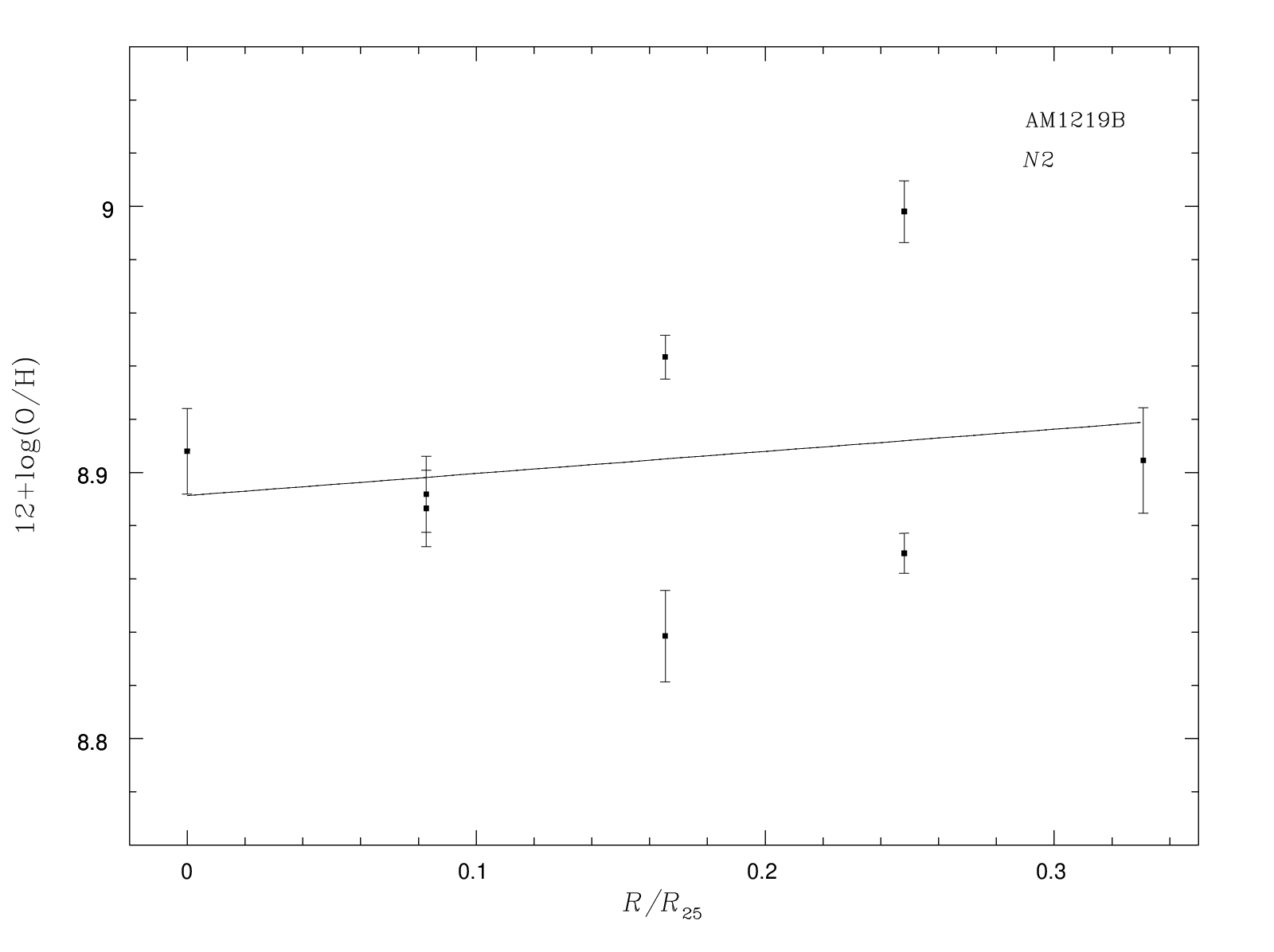}
\caption{Same as Fig.~\ref{grad_abu_1054B} but for  AM\,1219A and AM\,1219B. For AM\,1219A,  determinations
for the regions observed in different long-slit positions are indicated for distinct colours, as indicated. Also, linear regressions  
 considering all regions along the disk (dotted blue line) and inner and outer regions to galactocentric distance
 $R/R_{25}$=0.5  (solid black line) are shown.}
\label{grad_abu_1219AB}
\end{figure*}

In Fig.~\ref{grad_abu_1219AB}  the O/H distribution in both galaxies are shown.
 For AM\,1219B we derived a slope  $+0.10 \pm 0.18$  and a central oxygen abundance
of  $8.89 \pm 0.04$ dex. Such as for AM\,1054B, the slope for AM\,1219B was derived with few points  (and with a large dispersion), which does
the result highly uncertain. For AM\,1219A  the estimated oxygen gradient slopes are $-0.29\pm0.04$
  and $-0.54\pm 0.04$ using the $N2$ and $O3N2$ indexes, respectively, with a central   value of 12+log(O/H)$\sim 8.8$   derived from both indexes. 
 For values of $R/R_{25}$ from about 0.4 to  0.5, we can see a larger dispersion in the oxygen distribution
than the one found at other galactocentric distances. Regions at this
distance range, as can be seen in Fig.~\ref{fendas_1},
are located in the intersections of the  slits and are regions with high  surface brightness.
 It can be noted  in Fig.~\ref{grad_abu_1219AB}, for AM\,1219A,  that there is 
a change in the slope at  about $R/R_{25} =0.5$. The slopes of the abundance gradient in the inner disk ($R/R_{25} < 0.5\:$)
considering the   $O3N2$  and $N2$ indexes are  $-0.64\pm0.05$
and $-0.21 \pm0.05$, respectively. For the outer region ($R/R_{25} > 0.5\:$), the slopes are 
$+0.20\pm0.11$ and  $+0.16\pm0.11$ using $O3N2$  and $N2$, respectively.

\subsection{AM\,1256-433}

\begin{figure*}
 \includegraphics*[angle=270,width=0.8\textwidth]{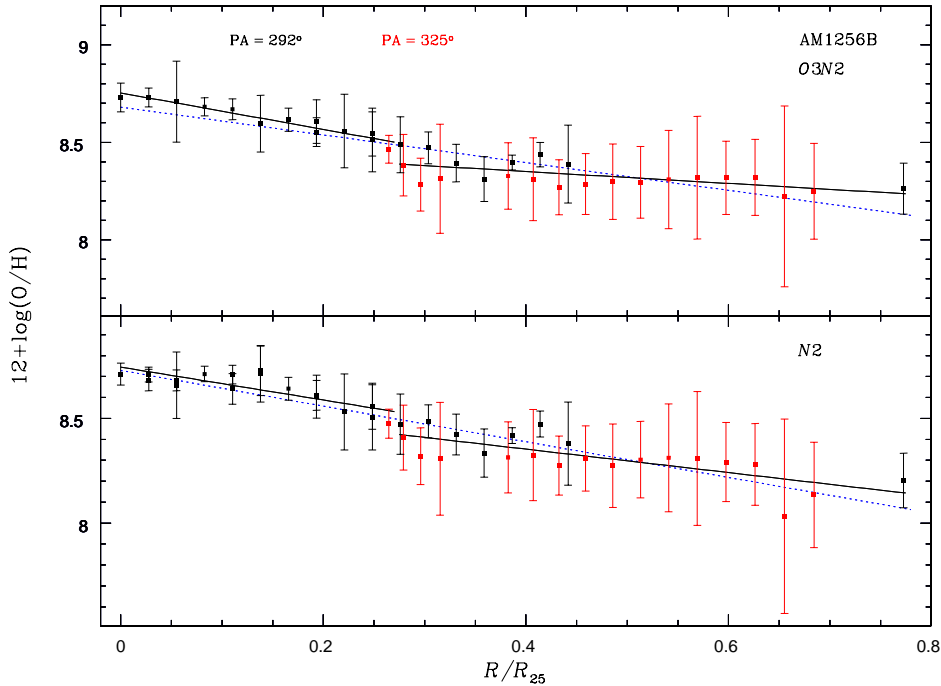}
\caption{Same as Fig.~\ref{grad_abu_1219AB} but for  AM\,1256B.  Linear regressions  
 considering all regions along the disk (dotted blue line) and inner and outer regions to galactocentric distance
 $R/R_{25}$=0.35 (solid black line) are shown.}
\label{grad_abu_1256B}
\end{figure*}

As was reported in Paper I, the AM\,1256-433 system is composed
by three galaxies  and  we only observed the AM\,1256-433B component.
For this object, the galactocentric distance measurements were corrected by inclination considering
 \textit{i}=77$^{\circ}$. In  Fig. \ref{grad_abu_1256B} the  O/H distribution
 via the $N2$
 and $O3N2$ indexes are shown. We obtained  gradient slopes  of $-0.85\pm0.06$ and $-0.71 \pm0.06$ for
 these indexes, respectively, with  a central 12+log(O/H) value  of $\sim 8.7$ dex.
Interestingly, we can note a steeper oxygen gradient for  $R/R_{25} \: < 0.27$ than the one 
obtained for the outer regions. The slopes of the abundance gradient in the inner disk
considering the  $N2$ and   $O3N2$ indexes are $-0.78\pm0.13$
and $-0.93 \pm0.07$. For the outer disk ($R/R_{25} > 0.27$)
we derived for  $N2$ and   $O3N2$ indexes the slopes 
$-0.55\pm0.10$ and $-0.30\pm0.08$, respectively.
The slopes of the global fits to the
abundance estimations are dominated by the values of the outer regions of the
galaxy, and they are slightly steeper than the ones obtained only considering
these outer regions.

\subsection{AM\,2030-303}
\begin{figure*}
 \includegraphics*[angle=270,width=0.8\textwidth]{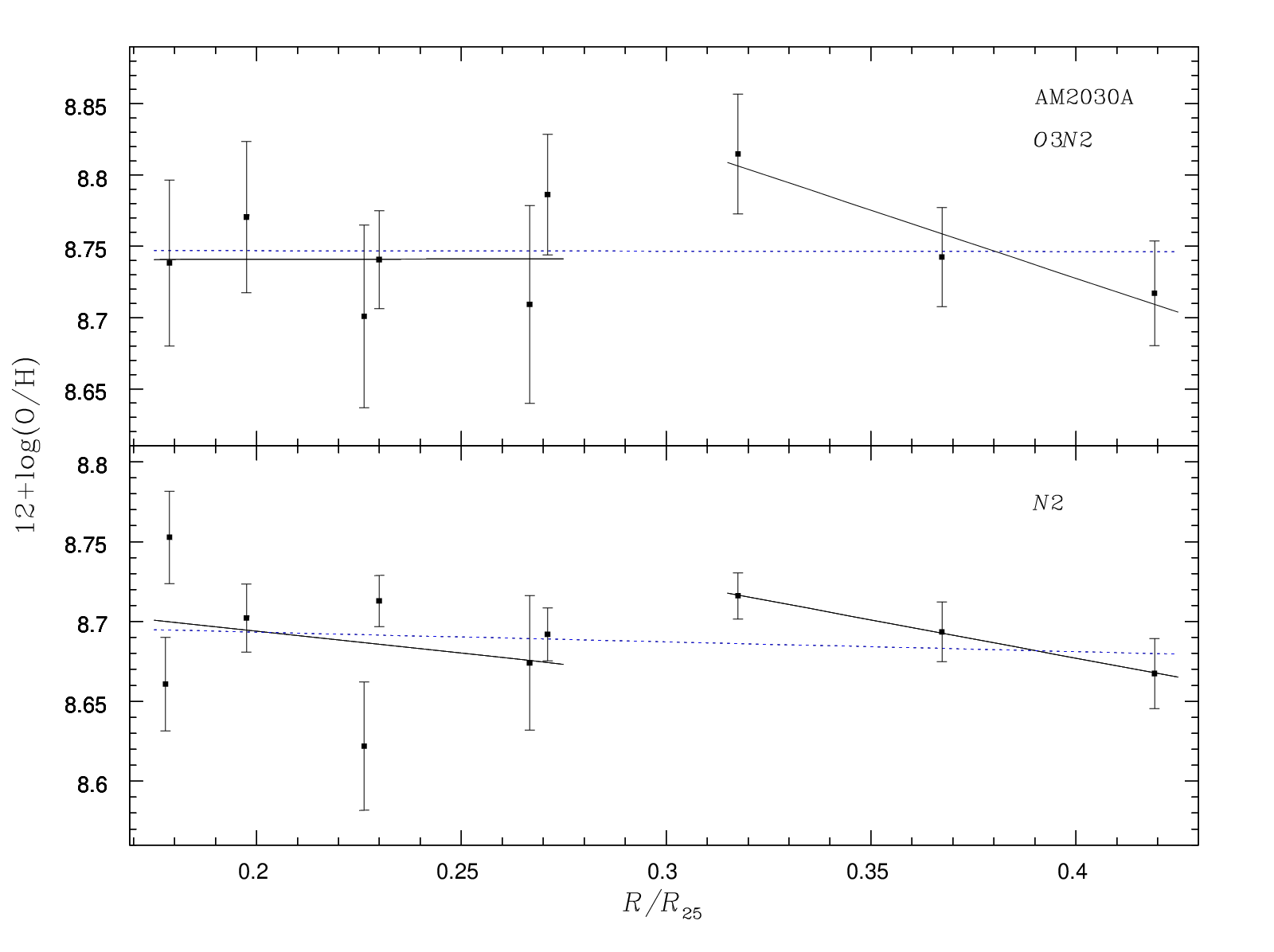}
 \includegraphics*[angle=270,width=0.8\textwidth]{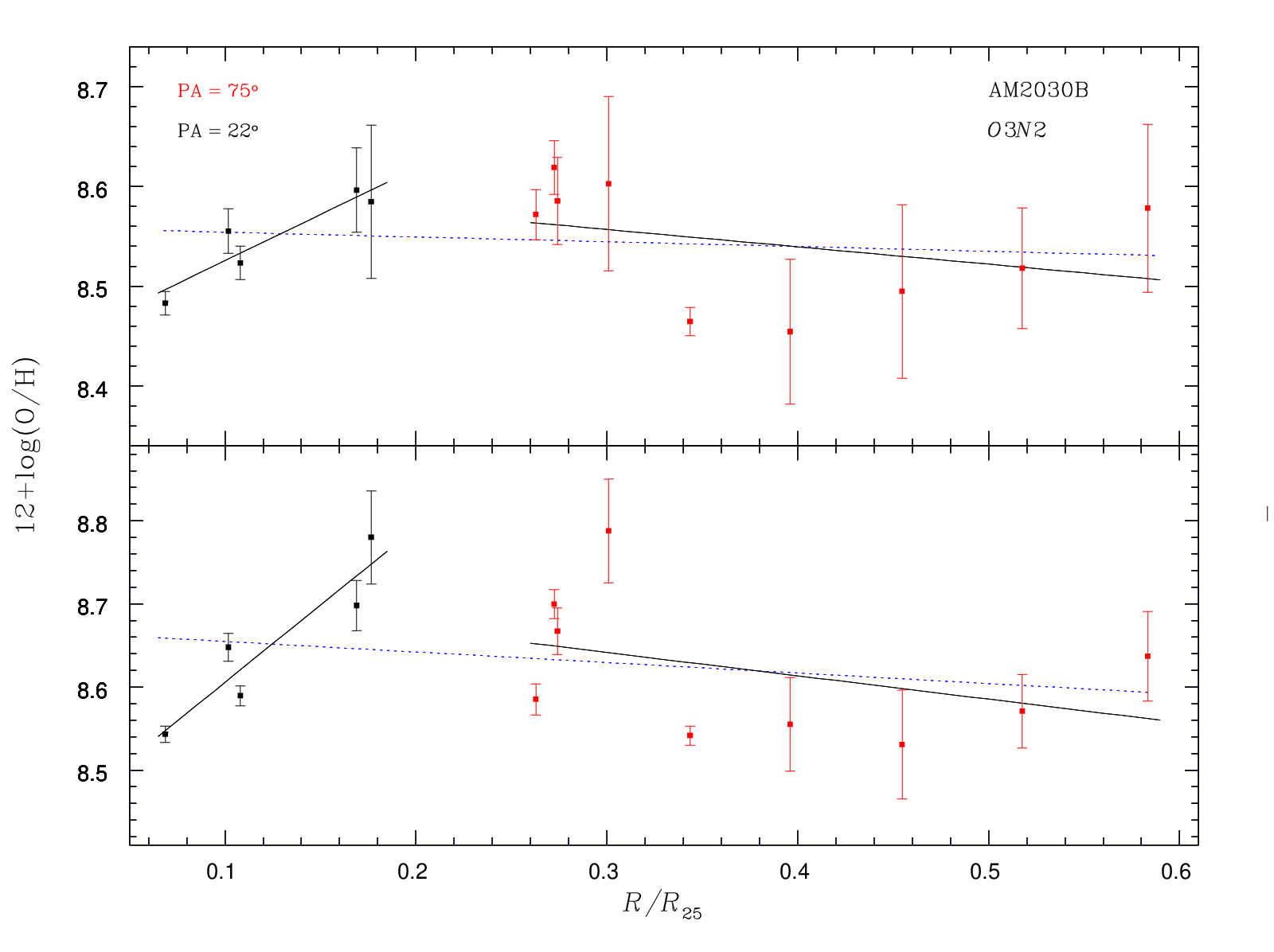}
\caption{Same as Fig.~\ref{grad_abu_1219AB} but for  AM\,2030A and AM\,2030B (ESO 463-IG 003 NED003).
  Linear  regressions  
 considering all regions along the disk (dotted blue line) and inner and outer regions to galactocentric distance
 $R/R_{25}$=0.2  (solid black line) are shown.}
\label{grad_abu_2030}
\end{figure*}

This system is composed by three galaxies, a main galaxy AM\,2030A (ESO 463-IG 003 NED01),  
the ESO 463-IG 003 NED02 and ESO 463-IG 003 NED03. These last two objects are 
a sub-system namely AM\,2030B. In Fig.~\ref{fendas_1} the  slit positions for each object is shown. 
We corrected by inclination the galactocentric distances of  AM\,2030B considering \textit{i}=35$^{\circ}$.
 Due to the few number of \ion{H}{ii} regions observed, it was not possible
to compute the O/H gradient for ESO 463-IG 003 NED002. In Fig.~\ref{grad_abu_2030}
 the results for  AM\,2030A and B are presented.
  
For AM\,2030A, considering a global fits,  we obtained a central value  12+log(O/H)=8.73$\pm$0.10   with 
a slope  $-0.31\pm$0.30 using $N2$ and  8.62$\pm$0.07 and 
$-0.22\pm$0.20 using $O3N2$.  These  values  
are similar to the highest O/H value obtained for the  \ion{H}{ii}  region CDT1 in NGC\,1232
by \citet{castellanos02}. Now,  considering an abundance gradient break,
the slopes  for  inner disk ($R/R_{25} < 0.27$)
obtained via $N2$ and   $O3N2$ indexes are $-0.27\pm0.46$
and $-0.95 \pm0.27$, respectively. For the outer disk ($R/R_{25} > 0.27$)
we derived for  $N2$ and   $O3N2$ indexes the slopes 
$-0.47\pm0.01$ and $0.00 \pm0.45$, respectively.

For AM\,2030B (ESO 463-IG 003 NED003)  we found 
 two  different  O/H abundance distributions at the inner and the outer regions of $R/R_{25}$=0.2.
The slopes of the abundance gradient in the inner disk ($R/R_{25} < 0.17$)
considering the  $N2$ and   $O3N2$ indexes are $+1.85\pm0.43$
and $+0.92 \pm0.22$, respectively. For the outer disk ($R/R_{25} > 0.17$)
we derived for  $N2$ and   $O3N2$ indexes the slopes 
$-0.28\pm0.25$ and $-0.17 \pm0.18$, respectively.

\subsection{AM\,2058-381}

\begin{figure*}
 \includegraphics*[angle=270,width=0.8\textwidth]{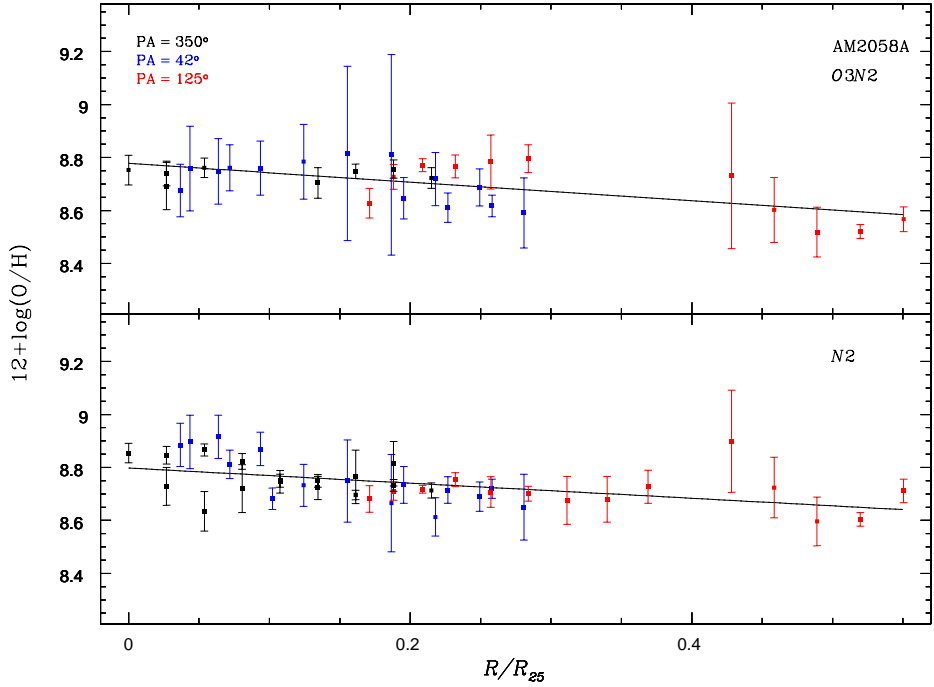}
\caption{Same as Fig.~\ref{grad_abu_1219AB}  but for AM\,2058A.}
\label{grad_abu_2058}
\end{figure*}

For this system the  O/H gradient   was only determined for one galaxy, 
AM\,2058A, since   the \ion{H}{ii} regions of its companion galaxy, AM\,2058B, 
 have emission lines   excited by gas shock (see Paper I). In Fig. \ref{grad_abu_2058} the oxygen
distribution is shown, where the galactocentric
distances  were  corrected considering \textit{i}=\,68$^{\circ}$.  We found 
a slope for the  O/H gradient  from the $N2$  index of
$-0.29\pm0.08$   
with  12+log(O/H)=8.79$\pm 0.02$ for the central region. Using the
$O3N2$, we 
found   $-0.35\pm0.08$  and  8.78$\pm 0.02$  dex.

\subsection{AM\,2229-735}

\begin{figure*}
 \includegraphics*[angle=270,width=0.8\textwidth]{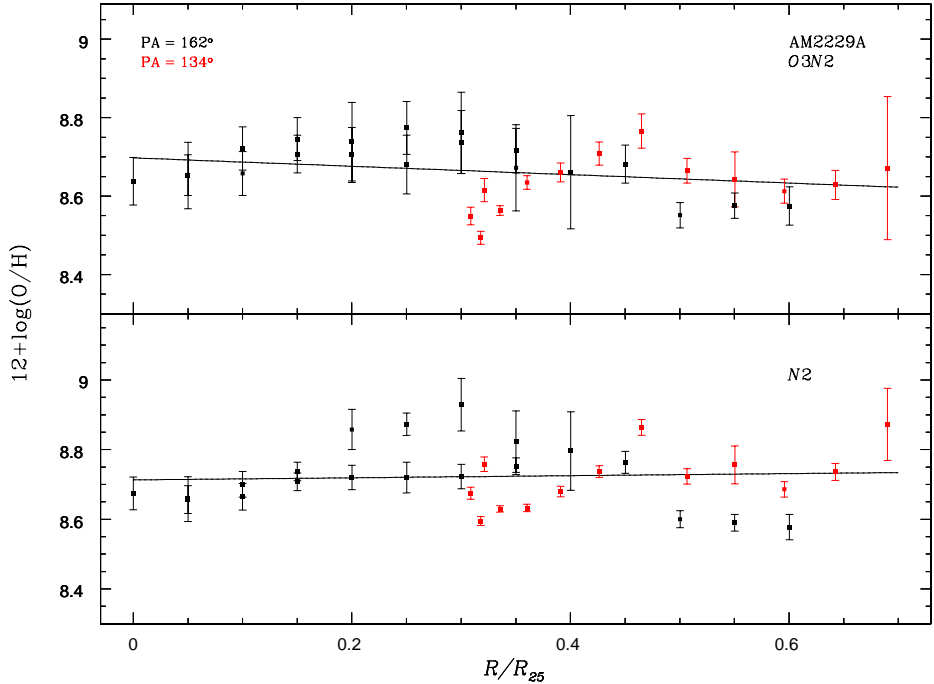}
\caption{Same as Fig.~\ref{grad_abu_1219AB}  but for AM\,2229A.}
\label{grad_abu_2229}
\end{figure*}

We obtained the  O/H abundance distributions only for the main galaxy of the
system AM\,2229-735, namely AM\,2229A and these are shown in
Fig. \ref{grad_abu_2229}. Galactocentric distances in this
  object were corrected by inclination considering 
\textit{i}=48$^{\circ}$. For estimations via the $N2$ index we found a slope of  
of  0.03$\pm$0.09   with  central oxygen abundance  12+log(O/H)=8.71$\pm$0.03 dex.
 For the $O3N2$ index these  values are $-0.11\pm$0.06 and 8.70$\pm$0.02 dex.

\subsection{AM\,2306-721}
\begin{figure*}
 \includegraphics*[angle=270,width=0.8\textwidth]{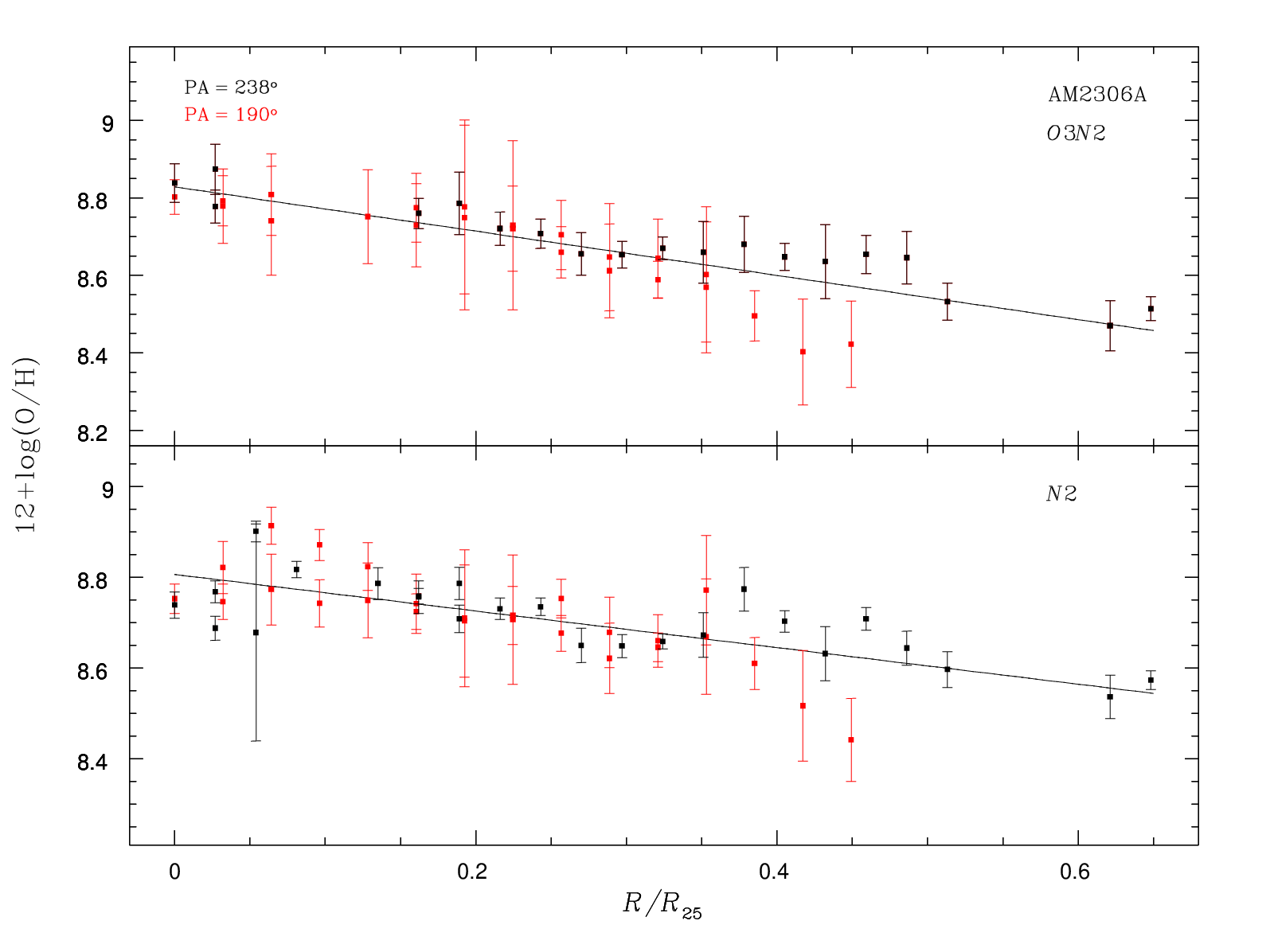}
\caption{Same as Fig.~\ref{grad_abu_1219AB}  but for  AM\,2306A.}
\label{grad_abu_2306}
\end{figure*}

This pair is composed by a main galaxy AM\,2306A e a companion galaxy AM\,2306B. 
The  O/H gradient (see Fig. \ref{grad_abu_2306}) was determined only for the  
main galaxy AM\,2306A because, for its companion, AM\,2306B, the presence of gas shock excitation along the disk was found (see Paper I).
 We found by using the $N2$  index a  slope  of $-0.40\pm$0.05  and
 a central value
 12+log(O/H)=8.81$\pm0.02$ dex. Using the $O3N2$ index the values $-0.57\pm$0.06   and  8.83$\pm0.02$ dex
 were found.

\subsection{AM\,2322-821}

\begin{figure*}
 \includegraphics*[angle=270,width=0.8\textwidth]{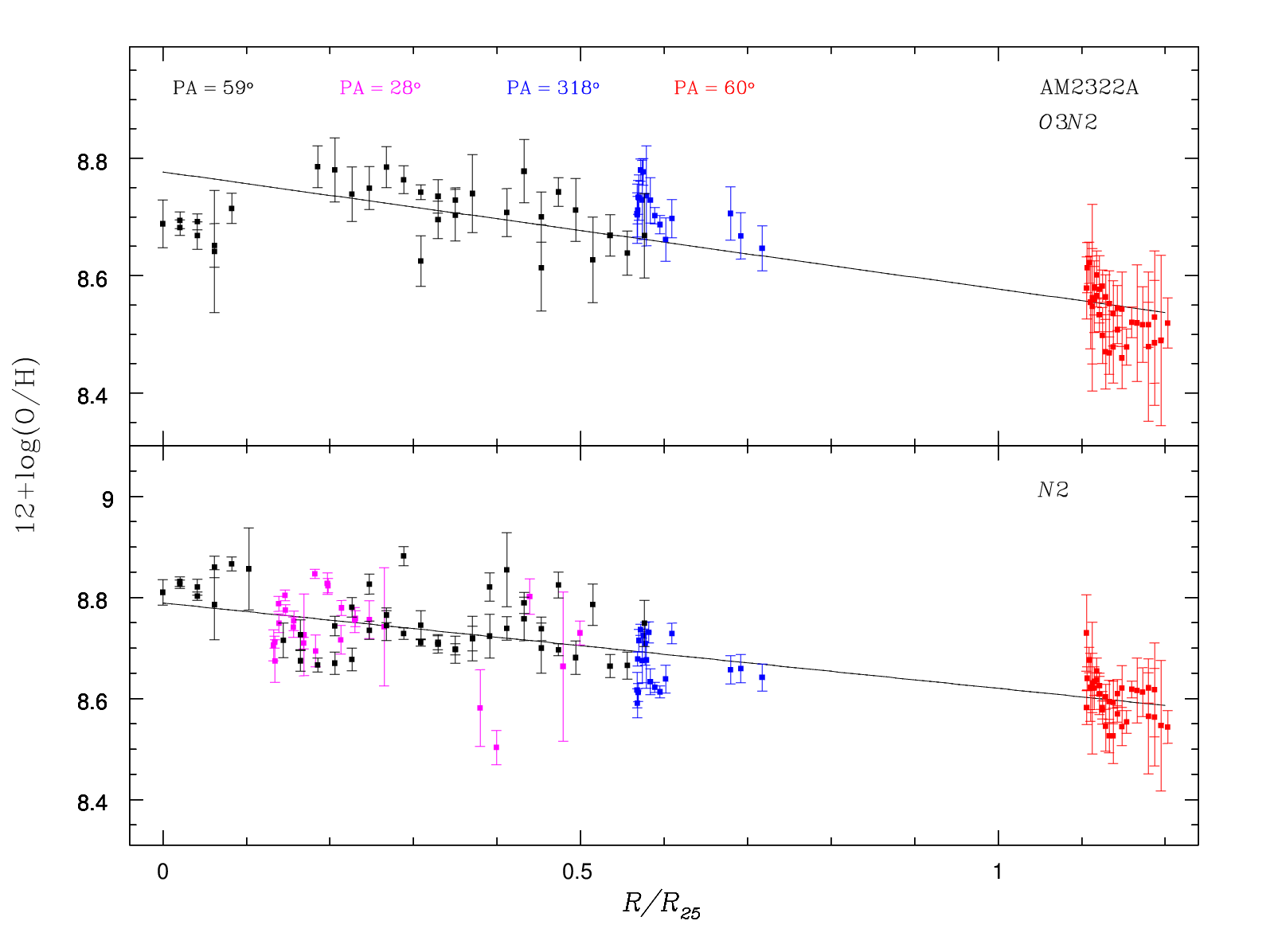}
 \includegraphics*[angle=270,width=0.8\textwidth]{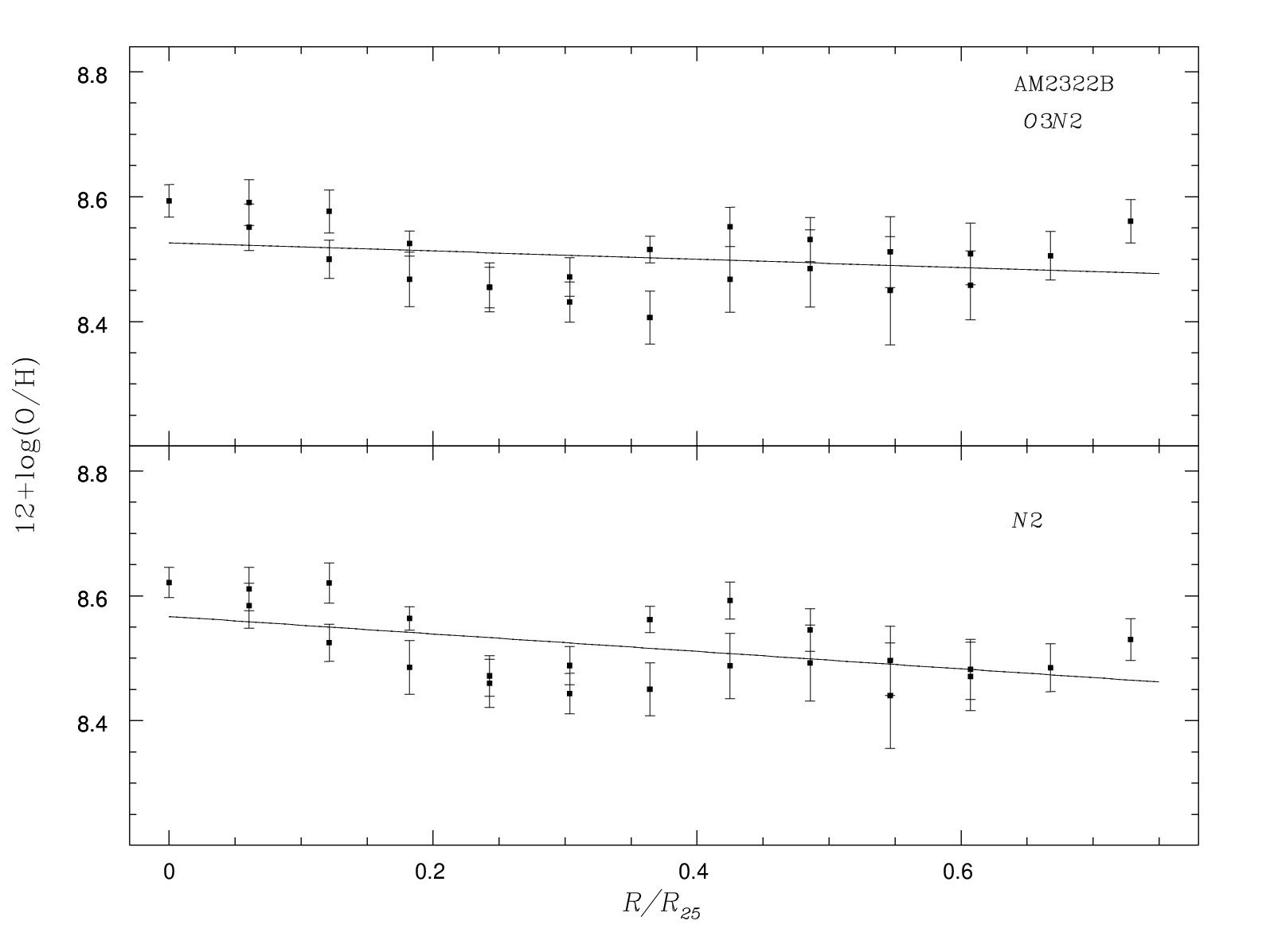}
\caption{Same as Fig.~\ref{grad_abu_1219AB} but for AM\,2322A and AM\,2322B.}
\label{grad_abu_2322}
\end{figure*}

In Fig.~\ref{grad_abu_2322}   the  O/H distributions along the disk of the pair of galaxies AM\,2322A and AM\,2322B
are shown. For AM\,2322A a correction for inclination was performed considering  \textit{i}=20$^{\circ}$.
For  AM\,2322A we obtained by using the $N2$ index a   slope of $-0.17\pm0.01$  and a  
12+log(O/H)=8.79$\pm0.01$ dex for the central part. From the $O3N2$ index, these values were found to be
 $-0.19\pm0.01$ and   8.77$\pm0.01$ dex. For the secondary object AM\,2322B,
the use of the $N2$ yielded  $-0.14\pm0.05$  and  8.57$\pm0.20$ dex for
the central part, while the use of the
 $O3N2$ yielded $-0.07\pm0.05$ and  8.53$\pm 0.20$ dex.

\section{Discussion}
\label{discussion}

Determinations of the oxygen  abundance gradients in interacting galaxies  have   showed that
these are shallower than the ones in isolated galaxies. 
This result was  obtained recently and 
few  works  have addressed this subject. 
In what follow some works in this directions are summarized.

\begin{enumerate}
\item \citet{krabbe08} obtained long slit spectroscopy data with the Gemini/GMOS
of the two components (A and B) of the galaxy pair AM\,2306-721. 
These authors used  a comparison
 between  the observed emission-line ratio intensities  ($R_{23}$ and
 [\ion{O}{ii}]$\lambda$3727/[\ion{O}{iii}]$\lambda$5007; see
 \citealt{mcGaugh91}) and those predicted by photoionization models 
 to determine the metallicity gradient.
 Krabbe and collaborators found a clear gradient for the most
   massive object of the pair, AM\,2306A, while for AM\,2306B 
 an oxygen abundance relatively homogeneous across the disc was found.

 \item \citet{lisa10}  selected 5 sets of close pairs  with separation between 15-25 kpc
from the sample of \citet{barton00} and obtained spectra for 12-40 star-forming regions in
8 of the close pair galaxies with the Keck Low-Resolution Imaging Spectrograph.  
This work was the first systematic investigation about  metallicity gradients in   close pairs. 
 \citet{lisa10}  found that the metallicity gradients in their sample are significantly shallower than gradients in isolated spiral galaxies.
 They used a theoretical calibration between the
   [\ion{N}{ii}]$\lambda$6584/[\ion{O}{ii}]$\lambda$3727 emission-line ratio
 and the metallicity.

\item \citet{krabbe11} obtained spectroscopic data of the two components (A and B) of the  system AM\,2322-821  with 
the Gemini/GMOS. The oxygen gradient was derived following the same procedure than in \citet{krabbe08}  
and a flat oxygen gradient was derived for  both  galaxies.  

\item  \citet{bresolin12}, who used the Focal Optical Reducer and Spectrograph (FORS2) attached
to the  Very Large Telescope,  obtained optical spectroscopy of  \ion{H}{ii} regions 
 belonging to  NGC\,1512. This galaxy has a companion, NGC\,1510,   separated by about 13.8 kpc.
 A   flatten radial abundance gradient   was also obtained  by using  several  
  calibrations  based on strong lines,  as well as some   oxygen
  abundance determinations using the  $T_{e}$-method.
 
\item  \citet{flores14} obtained Gemini/GMOS spectroscopic data  of the
interacting galaxy NGC\,92, which is part of a compact group and displays an extended
tidal tail. \citet{flores14} used  calibrations of the $N2$,  $O3N2$ and  [\ion{Ar}{iii}]$\lambda$7136/[\ion{O}{iii}]$\lambda$5007
indexes proposed by   \citet{pettini04} and \citet{stasinska06} to
 estimate the O/H abundance.  Torres-Flores and collaborators found
 that  most of the regions in NGC\,92 present a similar oxygen abundance, which produces an almost flat
metallicity gradient,  with a possible break in this gradient for the galactocentric distance of  $\sim10$ kpc.

\item  Most recently, \citet{sanchez14}, based on the largest and better defined statistical sample of galaxies yielded
by the CALIFA survey, compared the O/H abundance gradient of about  300 nearby galaxies, with more than 40
mergers/interacting systems, being half of them galaxy pairs.  \citet{sanchez14}  found, for the first time, a clear 
statistical evidence of a flattening in the abundance gradients in the interacting systems at any interaction stage, in agreement
with the previous results.

 \end{enumerate}
  
\begin{figure}
 \includegraphics*[angle=270,width=0.9\textwidth]{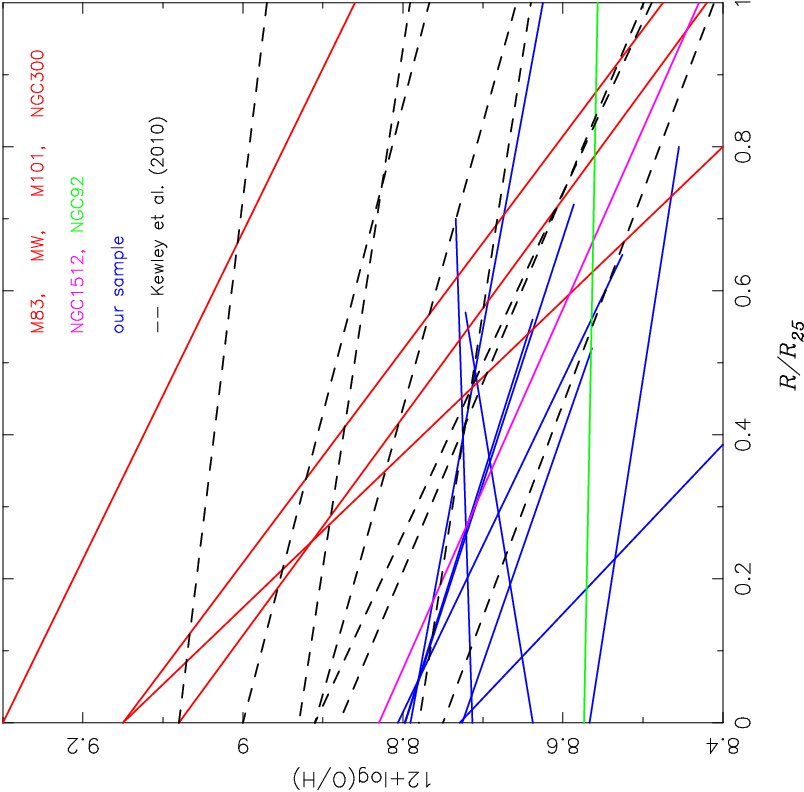}
\caption{Metallicity gradients from $N2$ for our sample.  The gradients for
AM\,1219B and AM\,1054B, derived with few points (see text), are not shown.
For comparison, we show the metallicity gradients for 
the isolated galaxies M\,101, Milk Way, M\,83, NGC\,300, whose the data were taken
from \citet{kennicutt03},  \citet{shaver83}, \citet{bresolin05}, and \citet{bresolin09}, respectively; 
and  eight interacting galaxies 
presented  by \citet{lisa10}, NGC\,92 \citep{flores14}  and  NGC\,1512 \citep{bresolin12}.}
\label{isolados}
\end{figure}

 From the literature summarized above it can be seen that  oxygen
 gradients   have been determined for close pairs or mergers systems 
 mainly  using different strong emission-line calibrations and 
 some few determinations using the $T_{e}$-method. In this paper, we
 performed a new determination of the O/H gradients 
for the galaxies  AM\,2306A  and  AM\,2322A-B previously studied by \citet{krabbe08, krabbe11} and presented 
an analysis of eight more galaxies in close pairs.  
In Fig.~\ref{isolados} the oxygen gradients from $N2$ derived for our
sample, for the 
galaxies in the pairs from the literature cited above, and 
for four  isolated ones, also from the literature, are
shown.  We can see that the oxygen gradients
derived for the objects in our sample
are shallower than the ones in isolated spirals. Although most of the slopes of these gradients are in
consonance with the ones obtained by  \citet{lisa10}, we found lower  O/H
abundances for the central parts of the galaxies.
This is due to the oxygen abundance estimations via theoretical calibrations, such as the one used by \citet{lisa10},
yield higher va\-lues than the ones from calibrations based on oxygen estimations via
$T_{e}$-method (see, for example, \citealt{dors05}). 
\citet{krabbe08}, \citet{lisa10}  and most recently, \citet{sanchez14},  interpreted the absence of abundance gradient  
in interacting galaxies as being due to the
 mixing produced by  low-metallicity gas from
the outer parts with the metal-rich gas of the centre of the galaxy.
 Here we confirmed this result by increasing the sample of objects.

 As we pointed out in Sect.~\ref{res3}, a flattening in the
oxygen gradient was found in the  outer  part of AM\,1256B  from $R/R_{25} \approx 0.35$ ($R \approx 8.5$ kpc), 
AM\,1219A from $R/R_{25}\approx 0.5$ ($R \: \approx \:7.6$ kpc) and  AM\,2030B
from $R/R_{25}\approx 0.2$ ($R \: \approx \:2.7$ kpc). In the case
  of AM\,1219A and AM\,1256B the inner gradients have negative slopes
  while AM\,2030B presents a positive inner gradient. In contrast,
  AM\,2030A has a positive-slope outer gradient while the inner one is
almost compatible with a flat behaviour, with the break at about
R/R$_{25} \approx 0.3$ ($R \approx 5$ kpc). However, if we take into
account the errors in the measurements, in the latter
case the slope of the outer gradient is also compatible with zero,
although due to the low number of \ion{H}{ii} regions in the outer
zone we are not able to give a conclusion.
The flattening in the oxygen gradients in the outer part  were also found  for either individual galaxies
\citep{rosales11, flores14,  bresolin12,  marino12, miralles14, martin95, zahid11, bresolin09, 
goddard11,marina14},  a small sample of interacting  galaxies \citep{werk11}, 
a large sample of objects  \citep{sanchez14, sanchez12a},  or even for the Milky Way  \citep{esteban13}.  
 Basically, there are four theoretical  scenarios to explain the flattening of the oxygen abundance gradients 
at a given galactocentric distance. (i)  The  pumping out effect of corotation, which produces gas flows in opposite
directions on the two sides of the resonance, yielding  
a minimum metallicity \citep{scarano13} and SFR \citep{mishurov02}.
(ii) A decrease of the star-formation efficiency as proposed by \citet{esteban13}. (iii) The accretion
of pristine gas \citep{marino12, sanchez14}. (iv) The bar presence (e.g. \citealt{zaritsky94, martin94}). 
 The data in our sample only allow us to investigate  the bar presence
and the star formation rate along the galactic disks. Inspection in the GMOS-S $r'$ acquisition images
of  AM\,1256B, AM\,1219A,  AM\,2030A  and AM\,2030B do not reveal the presence of any bar. Moreover, \citet{sanchez14},
investigated the effects of bars in the abundance gradients for the objects observed in the CALIFA survey.
\citet{sanchez14} did not found  differences in statistical terms  
between the slope of the abundance gradient for barred galaxies  and the 
one for other objects. Therefore, we excluded  the bar presence
  as the explanation for the flattening found in these four interacting 
galaxies of our sample.

To investigate if there is a minimum of SFR
at the break region, we used the $\rm H\alpha$ flux measured in our observation
and  the relation given by \citet{kennicutt98}
\begin{equation}
\label{sfr3}
 {\rm SFR}(M_{\odot}/{\rm yr})= 7.9  \times 10^{-42} \: L(\rm H\alpha) {\rm (erg/s)}.
\end{equation}

\begin{figure}
\includegraphics*[angle=270,width=0.9\textwidth]{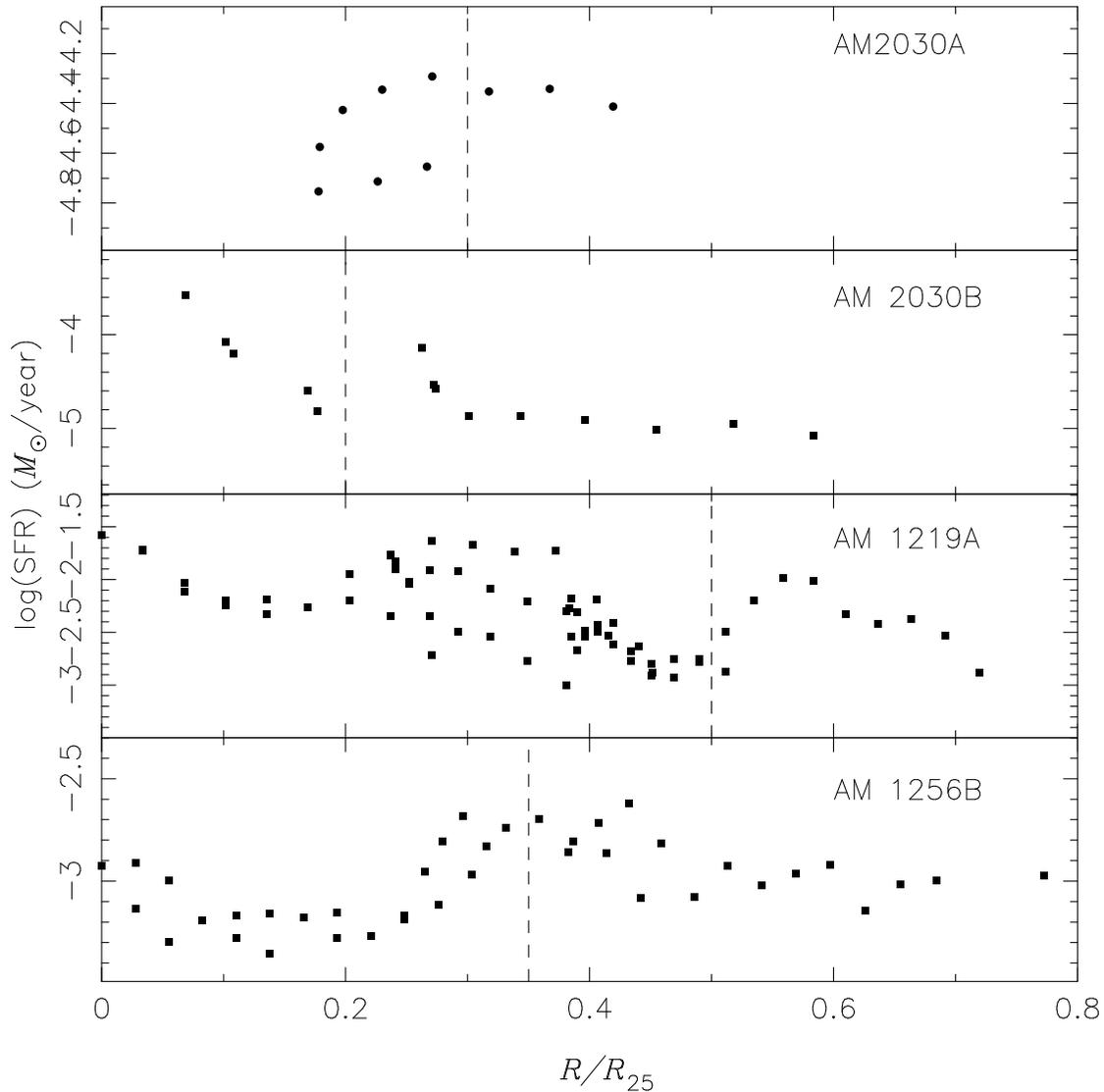}
\caption{SFR vs.\ the galactocentric distance ($R/R_{25}$) for AM\,1256B, 
AM\,1219A,  AM\,2030B and AM\,2030A.
The dotted line separates the  inner region where the steeper gradient
was found (see text).}
\label{sfr}
\end{figure}

Since the absolute flux of $\rm H\alpha$ was not obtained, our SFR values must be interpreted as a relative estimation
and the present analysis is only useful to study the behavior of SFR
along  the AM\,1256B, AM\,1219A,  AM\,2030A  and   AM\,2030B disks.
In Fig.~\ref{sfr} the SFR  versus $R/R_{25}$ for the galaxies above are shown, the inner region where the steeper gradient was found 
 is indicated.  For two objects, AM\,1219A and AM\,2030B,
     we can see that the SFR minimum values are located
 very close to the regions where the oxygen gradient breaks, in
 agreement with \citet{esteban13}. Only for
 the former we also found a minimum in the estimated metallicities indicating that this break zone
 could be associated with a corotation radius, as pointed out by
   \citet{mishurov02}. 
For the other two objects, AM\,1256B
   and AM\,2030A, the breaks in the abundance gradients are located
   very close to the SFR maximum.
 
\begin{figure*}
\includegraphics*[angle=270,width=0.9\textwidth]{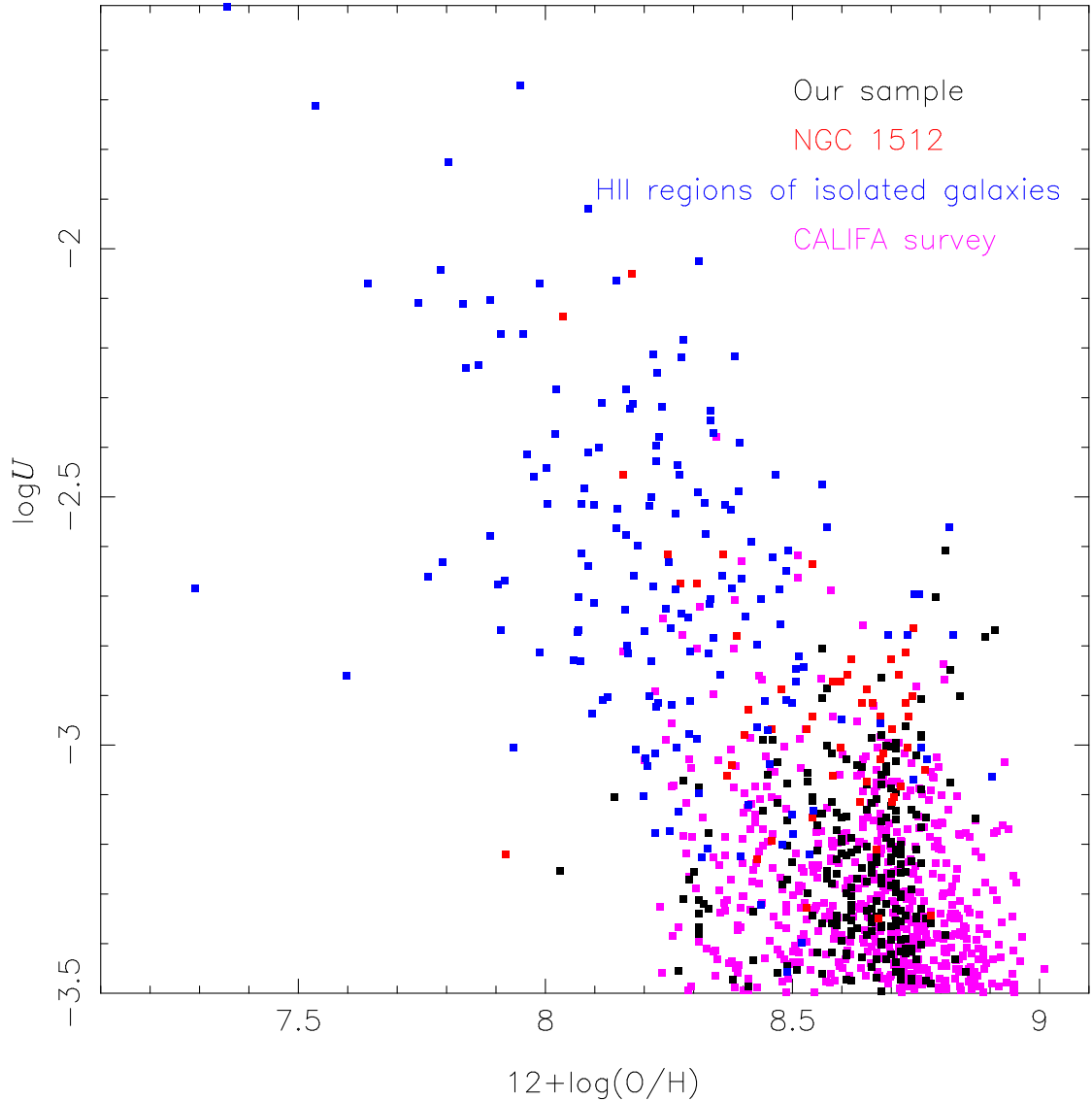}
\caption{Ionization parameter $U$ vs.\ the oxygen abundance of 
\ion{H}{ii} regions belonging to spiral galaxies from the CALIFA
survey, isolated galaxies  (data taken from \citealt{dors11}), 
NGC\,1512 \citep{bresolin12}, and our sample as indicated. 
Estimations of O/H  and $U$ were obtained using Eqs.~\ref{n2c} and \ref{ions2}, respectively.}
\label{ion}
\end{figure*}

Another important issue is the behavior of the ionization parameter $U$ with the metallicity.
 Basically, $U$ represents the dimensionless ratio of the ionizing photon density to the electron
density   and it is defined as $U=Q_{ion}/4\pi R^{2}_{\rm in} n  c$, where $Q_{ion}$  is the number of hydrogen ionizing photons emitted per second
by the ionizing source, $R_{\rm in}$  is  the distance from the ionization source to the inner surface
of the ionized gas cloud (in cm), $n$ is the  particle density (in $\rm cm^{-3}$), and $c$ is the speed of light.
Therefore, due to the gas flow along the disk of interacting galaxies yield high values of electron density when 
compared to the ones found in isolated star-forming regions (Paper I), it is
expected to find low $U$ values in the \ion{H}{ii}
regions located in our sample. To verify  that, we used the spectroscopic data presented in Table~\ref{tcorr1} and 
the relation  
\begin{equation}
\label{ions2}
\log U=-1.66\:(\pm0.06) \times  S2 -4.13\:(\pm 0.07),
\end{equation}
taken from \citet{dors11}, where $S2= \log([\ion{S}{ii}](\lambda6717,\lambda6731)/\rm H\alpha)$.
  This equation is valid for  $-1.5 \: \la \: \log U \: \la \: -3.5$ and estimations out of this range were not considered.
 These $U$ estimations are plotted in Fig.~\ref{ion} versus the O/H abundances determined from
$N2$ for our sample, for \ion{H}{ii} regions in the interacting galaxy NGC\,1512 observed  by \citet{bresolin12}
 as well as estimations for star-forming regions in  isolated galaxies obtained using the same 
calibrations and the data compiled by \citet{dors11}.   Also,   the CALIFA data \citep{sanchez12a}
for about  300  galaxies of any morphological type are included in this analysis. 
It can be seen that  \ion{H}{ii} regions located in interacting galaxies   do not present
the lowest $U$ values.  However, there is  a clear  correlation indicating  that
the  highest  abundances are found in those regions of galaxies with lower ionization strength
(see also \citealt{priscila13, enrique14}). In fact, \ion{H}{ii} regions located in the center of galaxies are
more evolved  (from their H$\alpha$ equivalent with) than the ones located 
in the outersticks regions, as pointed by \citet{sanchez12a},
having lower ionization strengths and higher, respectively.

\section{Conclusions}
\label{conc} 
   We presented an observational study about the  oxygen gradient abundance in 
   interacting galaxies. Long-slit spectra  in the range 4400-7300 \AA{}
were obtained  with the Gemini Multi-Object Spectrograph at Gemini South (GMOS)
for eleven galaxies  in eight close pairs.  Spatial profiles of oxygen abundance (used as metallicity tracer) in the gaseous phase along galaxy disks
were obtained   using calibrations based on strong emission-lines ($N2$  and $O3N2$). 
We found oxygen gradients 
significantly flatter for all galaxies in the close pairs of our
sample than the ones found in  isolated spiral galaxies. 
For four objects of
our sample, AM\,1219A, AM\,1256B, AM\, 2030A and
  AM\,2030B we found a clear break in the  oxygen abundance at
  galactocentric distances $R/R_{25}$  of about $0.5$, $0.35$, $0.3$, $0.2$, respectively. 
For two objects, AM\,1219A and AM\,1256B, we found negative slopes for
the inner gradients, and for AM\,2030B we found a positive one. In all
these three cases they   
show a flatter behaviour to the outskirts of the galaxies.
In the case of AM\,2030A, we found a positive-slope outer gradient while the inner one is
almost compatible with a flat behaviour.
This result is not concluding due to the small number of measured \ion{H}{ii}
regions mainly for the outer part.
We found a decrease of star formation efficiency in the zone that
corresponds to the oxygen abundance gradient break for AM\,1219A and
AM\,2030B. Moreover, in the case of AM\,1219A we also found a minimum
in the estimated metallicities indicating that this break zone could
be associated with a corotation radius. For the other two galaxies
that present a gradient break, AM\,1256B and AM\,2030A, we found a
maximum for the SFR but not an extreme oxygen abundance value. 
It must be noted that for all these four interacting systems the
extreme SFR values are located very close to the oxygen
gradient break zones.
The flattening in the oxygen abundance gradients could be interpreted  as being
a chemical enrichment due to
induced star formation by gas flows along the disks. 
We have found that  \ion{H}{ii} regions located in 
close pairs of galaxies 
follow the same relation between the
ionization parameter and the oxygen abundance as those regions in
isolated galaxies.

\section*{Acknowledgements}

Based on observations obtained at the Gemini Observatory, which is operated by 
the Association of Universities for Research in Astronomy, Inc., under a 
cooperative agreement with the NSF on behalf of the Gemini partnership: the 
National Science Foundation (United States), the Science and Technology
Facilities Council (United Kingdom), the National Research Council (Canada), CONICYT
(Chile), the Australian Research Council (Australia), 
Minist\'erio da Ciencia e Tecnologia (Brazil), and SECYT (Argentina).
D. A. Rosa, O. L. Dors Jr and A. C. Krabbe thanks the support of
FAPESP, process 2011/08202-6, 2009/14787-7 and  2010/01490-3,
respectively. 
We also thank to the anonymous referee for her/his careful and
constructive revision of this manuscript.


\begin{thebibliography}{99}
\bibitem[Alonso et al. (2012)]{alonso12} Alonso-Herrero, A.,  Rosales-Ortega, F.~F.,  S\'anchez, S.~F. et al. 2012,  MNRAS, 425, L46
\bibitem[Alloin et al.(1979)]{alloin79} Alloin, D., Collin-Souffrin, S., Joly, M., Vigroux, L., 1979, A\&A, 78, 200
\bibitem[Andrievsky et al.(2002)]{andrievsky02} Andrievsky, S.~M., Kovtyukh, V.~V., Luck, R.~E. et al. 2002, A\&A, 392, 491
\bibitem[Andrievsky et al.(2004)]{andrievsky04} Andrievsky, S.~M., Luck, R.~E., Martin, P., L\'epine, J.~R.~D. 2004, A\&A, 413, 159
\bibitem[Athanassoula(1992)]{athanassoula92}  Athanassoula, E. 1992, MNRAS, 259, 345
\bibitem[Barden et al.(2005)]{barden05} Barden, M., Rix, H.-W., Somerville, R.S. et al. 2005, ApJ, 635, 959
\bibitem[Barton et al.(2000)]{barton00} Barton, E.~J., Geller, M.~J.,  Kenyon, S.~J. 2000, ApJ, 530, 660 
\bibitem[Bastian et al.(2009)]{bastian09} Bastian, N., Trancho,  G., Konstantopoulos I.~S., Miller, B.~W., 2009, ApJ, 701, 607 
\bibitem[Bernloehr(1993)]{bernloehr93}  Bernloehr, K. 1993, A\&A, 270, 20
\bibitem[Bergvall et al.(2003)]{bergvall03}  Bergvall, N.,  Laurikainen, E.,  Aalto, S. 2003, A\&A, 405, 31
\bibitem[Bragaglia et al. (2008)]{bragaglia08} Bragaglia, A., Sestito, P., Villanova, S. et al. 2008, A\&A, 480, 79
\bibitem[Bresolin et al.(2012)]{bresolin12} Bresolin, F., Kennicutt R. C., Ryan-Weber E., 2012, ApJ, 750, 122
\bibitem[Bresolin(2011)]{bresolin11} Bresolin, F. 2011, ApJ, 730, 129
\bibitem[Bresolin et al.(2009)]{bresolin09} Bresolin, F.,  Gieren, W., Kudritzki, R.~-P. et al.  2009, ApJ, 700, 309
\bibitem[Bresolin et al.(2005)]{bresolin05} Bresolin, F., Schaerer, D., Gonz\'alez Delgado, R.~M., Stasi\'nska, G. 2005, A\&A, 441, 981
\bibitem[Boissier \& Prantzos(2000)]{boissier00} Boissier, S.,  \& Prantzos, N. 2000, MNRAS, 312, 398
\bibitem[Bell \& Jong(2000)]{bell00}  Bell, E.~F., de Jong, R.~S. 2000, MNRAS, 312, 497
\bibitem[Bresolin et al. (2009)]{bresolin09} Bresolin, F., Ryan-Weber, E., Kennicutt, R.~C., Goddard, Q. 2009, ApJ, 695, 580
\bibitem[Cardelli et al.(1989)]{cardelli89} Cardelli J.~A., Clayton,  G.~C., Mathis J.~S., 1989, ApJ, 345, 245
\bibitem[Castellanos et al.(2002)]{castellanos02} Castellanos, M.,  D\'{\i}az, A.~I., Terlevich, E. 2002, MNRAS, 329, 315
\bibitem[Chien et al.(2007)]{chien07}  Chien, L.,  Barnes, J.~E., Kewley, L.~J., Chambers, K.~C., 2007, ApJ, 660, L105.
\bibitem[Cid Fernandes et al.(2005)]{cid05} Cid Fernandes, R., Mateus,  A., Sodr\'e,  L. et al. 2005, MNRAS, 358, 363
\bibitem[Costa et al.(2004)]{costa04} Costa, R.~D.~D.,  Uchida, M.~M.~M., Maciel, W.~J. 2004, A\&A, 423, 199
\bibitem[Dalcanton(2007)]{dalcanton} Dalcanton, J.~J. 2007, ApJ, 658, 941
\bibitem[Di Matteo et al.(2008)]{dimatteo08} Di Matteo, P.,  Bournaud, F.,  Martig, M. 2008, A\&A, 492, 31
\bibitem[Donzelli \& Pastoriza(1997)]{donzelli97} Donzelli, C.~J., Pastoriza, M.~G., 1997, ApJS, 111, 181
\bibitem[Dors et al.(2011)]{dors11}  Dors, O.~L., Krabbe, A.~C., H\"agele, G.~F., P\'erez-Montero, E. 2011, MNRAS, 415, 3616
\bibitem[Dors \& Copetti(2005)]{dors05}  Dors, O.~L.,  \& Copetti, M.~V.~F. 2005, A\&A, 437, 837
\bibitem[Esteban et al.(2013)]{esteban13} Esteban, C.,  Carigi, L.,  Copetti, M.~V.~F. et al. 2013, MNRAS, 433, 382
\bibitem[Ellison et al.(2013)]{sara13} Ellison S. L., Mendel J. T., Patton D. R., Scudder J. M., 2013, MNRAS, 435, 3627
\bibitem[Ferreiro \& Pastoriza(2004)]{ferreiro04} Ferreiro, D.~L., \& Pastoriza, M.~G. 2004, A\& A, 428, 837
\bibitem[Ferreiro et al.(2008)]{ferreiro08} Ferreiro, D.~L., Pastoriza, M.~G., Rickes, M., 2008, A\&A, 481, 645
\bibitem[Freedman Woods et al.(2010)]{woods10} Freedman Woods, D.,  Geller, M.~J.,  Kurtz, M.~J. et al. 2010, AJ, 139, 1857
\bibitem[Freitas-Lemes et al.(2013)]{priscila13}  Freitas-Lemes, P., Rodrigues, I,  F\'aundez-Abans, M.,  Dors, O.~L.,
Fernandes, I.~F. 2013, MNRAS, 427, 2772 
\bibitem[Friedli et al.(1994)]{friedli94} Friedli, D., Benz, W.,   Kennicutt, R.~C. 1994, ApJ, 430, L105
\bibitem[Garnett et al.(1997)]{garnett97} Garnett, D.~R., Shields, G.~A., Skillman, E.~D., Sagan, S.~P. , Dufour, R.~J. 1997, ApJ, 489, 63.
\bibitem[Goddard et al.(2011)]{goddard11} Goddard, Q.~E., Bresolin F., Kennicutt R.~C., Ryan-Weber, E.~V., Rosales-Ortega, F.~F. 2011, MNRAS, 412, 1246
\bibitem[H\"agele et al.(2008)]{hagele08} H\"agele,  G.~F., D\'{\i}az,  A.~I., Terlevich E. et al.  2008, MNRAS, 383, 209
\bibitem[Hernandez-Jimenez et al.(2013)] Hernandez-Jimenez, J.~A., Pastoriza,  M.~G., Rodrigues, I. et al. 2013, MNRAS, 435, 3342 
\bibitem[Hummer \& Storey(1987)]{hummer87}  Hummer, D.~G., \& Storey P.~J. 1987, MNRAS, 224, 801 
\bibitem[Hernandez-Jimenez et al.(2013)]{jose13} Hernandez-Jimenez, J.~A., Pastoriza, M.~G.,
Rodrigues, I., Krabbe, A.~C., Winge, C., Bonatto, C. 2013, MNRAS, 435, 3342	
\bibitem[L\'opez-S\'anchez \& Esteban(2010)]{lopez10} L\'opez-S\'anchez, A.~R., \& Esteban, C. 2010, A\&A, 516, 104
\bibitem[Luck et al.(2003)]{luck03} Luck, R.~E., Gieren, W.~P., Andrievsky, S.~M. et al. 2003, 401, 939
\bibitem[Lemasle et al.(2013)]{lemasle13} Lemasle, B., François, P., Genovali, K. et al. 2013, A\&A, 558, 31
\bibitem[Kennicutt et al.(2003)]{kennicutt03} Kennicutt, R.~C.,  Bresolin, F.,  Garnett, D.~R. 2003, ApJ, 591, 801
\bibitem[Kennicutt(1998)]{kennicutt98} Kennicutt, R.~C. 1998, ARAA, 36, 189
\bibitem[Kewley et al.(2010)]{lisa10} Kewley,  L.~J., Rupke,  D., Zahid,  H.~J., Geller,  M.~ J., Barton, E.~J., 2010, ApJ, 721, L48
\bibitem[Kewley \& Ellison(2008)]{lisa08} Kewley, L.~J.,  \& Ellison, S.~L., 2008, ApJ, 681, 1183
\bibitem[Krabbe et al.(2014)]{krabbe14}   Krabbe, A.~C., Rosa, D.~A., Dors,  O.~L. et al. 2014, MNRAS, 437, 1155
\bibitem[Krabbe et al.(2011)]{krabbe11}  Krabbe, A.~C., Pastoriza, M.~G., Winge, C. et al. 2011, MNRAS, 416, 38 
\bibitem[Krabbe et al.(2008)]{krabbe08} Krabbe, A.~C.,  Pastoriza, M.~G.,  Winge, C.,  Rodrigues, I., Ferreiro, D.~L. 2008,  MNRAS, 389, 1593
\bibitem[Maciel \& Costa(2009)]{maciel09} Maciel, W.~J., \& Costa, R.~D.~D. 2009, in IAU Symposium, Vol. 254, IAU
Symposium, ed. J. Andersen, Nordstr\"oara, B. m, \& J. Bland-Hawthorn, 38P
\bibitem[MacArthur et al.(2004)]{mac04} MacArthur, L.~A., Courteau, S., Bell, E., Holtzman, J.~A. 2004, ApJS, 152, 175
\bibitem[McGaugh(1991)]{mcGaugh91} McGaugh, S.~S. 1991, ApJ, 380, 140
\bibitem[Magrini et al. (2009)]{magrini09} Magrini, L., Sestito, P., Randich, S., Galli, D. 2009, A\&A, 494, 95
\bibitem[Marino et al.(2012)]{marino12} Marino, R.~A.,  Gil de Paz, A.,  Castillo-Morales, A. et al. 2012, ApJ, 754, 61
\bibitem[Marino et al.(2013)]{marino13} Marino, R.~A., Rosales-Ortega, F.~F., S\'anchez, S.~F. et al. 	2013, A\&A, 559, 114
\bibitem[Martin \& Roy(1994)]{martin95} Martin, P., \& Roy, J.~R. 1995, ApJ, 445, 161
\bibitem[Martin \& Roy(1994)]{martin94} Martin, P., \& Roy, J.~R. 1994, ApJ, 424, 599 
\bibitem[Miralles-Caballero et al.(2014)]{miralles14} Miralles-Caballero, D., D\'{\i}az, A.~I., Rosales-Ortega, F.~F., P\'erez-Montero, 
E., S\'anchez, S.~F. 2014, MNRAS, 440, 2265	
\bibitem[Mihos et al.(2010)]{mihos10} Mihos, J.~C.,  Bothun, G.~D.,  Richstone, D.~O. 2010,  ApJ, 418, 82
\bibitem[Mishurov et al.(2002)]{mishurov02} Mishurov, Y.~N., L\'epine, J.~R.~D., Acharova, I.~A., 2002, ApJ, 571, L113
\bibitem[M\'olla \& D\'{\i}az(2005)]{molla05} M\'olla,  M., \& D\'{\i}az, A.~I. 2005, MNRAS, 358, 521
\bibitem[Mu\~noz-Mateos et al.(2007)]{munoz07} Mu\~noz-Mateos, J.~C., Gil de Paz, A., Boissier, S. et al. 2007, ApJ, 658, 1006	
\bibitem[Nikolic et al.(2004)]{nikolic04}   Nikolic, B., Cullen, H., Alexander, P., 2004, MNRAS, 355, 874
\bibitem[Lambas et al.(2003)]{lambas03} Lambas, D.~G., Tissera, P.~B., Alonso, M.~S.,  Coldwell, G. 2003, MNRAS, 346, 1189
\bibitem[Patton et al.(2011)]{patton11} Patton, D.~R.,  Ellison, S.~L.,  Simard, L. et al. 2011, MNRAS, 412, 591
\bibitem[Paturel et al.(2003)]{paturel03}Paturel, G., Petit, C., Prugniel, P. et al. 2003, A\&A, 412, 45
\bibitem[Paturel et al.(1991)]{paturel91} Paturel, G.,  Garcia, A.~M., Fouque,  P., Buta,  R. 1991, A\&A, 243, 319
\bibitem[Pedicelli et al.(2009)]{pedicelli09} Pedicelli, S., Bono, G., Lemasle, B. et al. 2009, A\&A, 504, 81
\bibitem[P\'erez-Montero \& Contini(2009)]{perez09} P\'erez-Montero, E., \&  Contini, T., 2009, MNRAS, 398, 949
\bibitem[P\'erez-Montero(2014)]{enrique14} P\'erez-Montero, E. 2014, MNRAS, 441, 2663
\bibitem[Pettini \& Pagel(2004)]{pettini04}  Pettini M., \& Pagel B. E. J., 2004, MNRAS, 348, L59
\bibitem[Pilyugin et al.(2012)]{pilyugin12} Pilyugin, L.~S., Grebel, E.~K., Mattsson, L.  MNRAS, 424, 2316
\bibitem[Pilyugin(2003)]{pilyugin03} Pilyugin, L.~S.  2003, A\&A, 397, 109
\bibitem[Pilyugin \& Thuan(2005)]{pilyugin05} Pilyugin, L.~S., \& Thuan, T.~X. 2005, ApJ, 631, 231
\bibitem[Pohlen \& Trujillo(2006)]{pohlen06} Pohlen, M., \& Trujillo, I. 2006, A\&A, 454, 759
\bibitem[Portinari \& Chiosi(1999)]{portinari99} Portinari, L., \& Chiosi, C. 1999, A\&A, 350, 827
\bibitem[Rodr{\'{\i}}guez-Baras et al.(2014)]{marina14} Rodr{\'{\i}}guez-Baras, M., Rosales-Ortega, F.~F., D{\'{\i}}az, A.~I., S{\'a}nchez, S.~F., \& Pasquali, A.\ 2014, \mnras, 442, 495
\bibitem[Rosales et al. (2011)]{rosales11}  Rosales-Ortega, F.~F.,  D\'{\i}az,  A.~I.,  Kennicutt, R.~C., S\'anchez, S.~F. 2011, MNRAS, 415, 2439
\bibitem[Rupke et al.(2010a)]{rupke10a} Rupke, D.~S.~N., Kewley, L.~J., Barnes J. E.  2010a, ApJ,  710, L156
\bibitem[Rupke et al.(2010b)]{rupke10b} Rupke, D.~S.~N., Kewley, L.~J., Chien,  L.~-H. 2010b, ApJ, 723, 1255
\bibitem[Scarano \& L\'epine(2013)]{scarano13} Scarano, S., \& L\'epine,  J.~R.~D., 2013, MNRAS, 428, 625
\bibitem[Scarano et al.(2011)]{scarano11} Scarano, S.,  L\'epine,  J.~R.~D., Marcon-Uchida, M.~M., 2011, MNRAS, 412, 1741
\bibitem[S\'anchez et al.(2012)]{sanchez12a} S\'anchez, S.~F., Rosales-Ortega, F.~F., Marino, R. A. et al. 2012a, A\&A, 546, 2
\bibitem[S\'anchez et al.(2014)]{sanchez14} S\'anchez, S.~F., Rosales-Ortega, F.~F., Iglesias-P\'aramo, J. et al. 2014, A\&A, 563, 49
\bibitem[Scudder et al.(2012)]{scudder12} Scudder, J.~M., Ellison, S.~L., Torrey,  P., Patton, D.~R., Mendel,  J.~T., 2012, MNRAS, 426, 549
\bibitem[Shaver et al.(1983)]{shaver83} Shaver, P.~A., McGee, R.~X., Newton, L.~M., Danks, A.~C., Pottasch, S.~R. 1983, MNRAS, 204, 53
\bibitem[Stasi\'nska(2006)]{stasinska06} Stasi\'nska G., 2006, A\&A, 454, L127
\bibitem[Storchi-Bergmann  et al.(1994)]{thaisa94} Storchi-Bergmann T., Calzetti,  D., Kinney,  A.~L. 1994, ApJ, 429, 572
\bibitem[Toomre \& Toomre(1972)]{toomre72} Toomre, A.,  \& Toomre Juri 1972, ApJ, 178, 623
\bibitem[Torres-Flores et al.(2014)]{flores14} Torres-Flores, S., Scarano, S., Mendes de Oliveira, C., de Mello, D.~F., Amram, P., Plana, H., 2014, MNRAS, 438, 1894
\bibitem[Trancho et al.(2007)]{trancho07} Trancho, G., Bastian, N., Miller, B.~W., Schweizer, F. 2007, ApJ, 664, 284
\bibitem[Trujillo et al.(2004)]{trujillo04} Trujillo, I., Rudnick, G.,  Rix, H.-W. et al. 2004, ApJ, 604, 521
\bibitem[V\'{\i}lchez et al.(1996)]{vilchez96} V\'{\i}lchez,  J.~M., \& Esteban, C. 1996, MNRAS, 280, 720
\bibitem[Zahid \& Bresolin(2011)]{zahid11} Zahid, H.~J., \&  Bresolin,  F. 2011, AJ, 141, 192
\bibitem[Zaritsky et al.(1994)]{zaritsky94} Zaritsky, D., Kennicutt, R.~C.,  Huchra, J.~P. 1994, ApJ, 420, 87 
\bibitem[Werk et al.(2011)]{werk11}  Werk, J.~K., Putman, M.~E., Meurer, G.~R., Santiago-Figueroa,  N., 2011, ApJ, 735, 71
\bibitem[Wright(2006)]{2006PASP..118.1711W} Wright, E.~L.\ 2006, \pasp, 
118, 1711 
\bibitem[Yong et al.(2014)]{yong14} Yong, D., Carney, B.~W., Friel, E.~D. 2014, AJ, 144, 95
\end{thebibliography}
\end{document}